\documentclass[useAMS,usenatbib,fleqn]{mn2e}
\usepackage{amstext}
\usepackage{amsfonts}
\usepackage{amssymb}
\usepackage{amsmath}

\usepackage{graphicx}
\usepackage{caption}
\usepackage{epstopdf}
\usepackage[varg]{txfonts}
\usepackage{xspace}
\usepackage{url}
\usepackage{multirow}
\usepackage{multicol}        
\usepackage{bm}		
\usepackage{pdflscape}	
\usepackage[deletedmarkup=sout]{changes}
\definechangesauthor[]{wek}
\usepackage{color}

\usepackage{times}

\newcommand{\sncut}{20}
\newcommand{\bgcut}{0.8}	

\newcommand{\goodstar}{704} 
\newcommand{\goodstarmemberadd}{649}

\newcommand{\angstrom}{\mathring{A}}

\newcommand{\teff}{\ensuremath{T_\textrm{eff}}}

\newcommand{\logg}{\ensuremath{\log(g)}}
\newcommand{\mh}{[M/H]}
\newcommand{\kms}{km\,s$^{-1}$}
	\title[Asymmetric spatial distribution of sub-solar metallicity stars in the Milky Way nuclear star cluster]{Asymmetric spatial distribution of sub-solar metallicity stars in the Milky Way nuclear star cluster \thanks{Based on observations collected at the European Organisation for Astronomical Research in the Southern Hemisphere, Chile (60.A-9450(A), 091.B-0418, 093.B-0368, 195.B-0283.}}
	\author[A.~Feldmeier-Krause, et al.]
	{A.~Feldmeier-Krause,$^{1}$\thanks{E-mail: afeldmei@uchicago.edu} 
   W.~Kerzendorf,$^{2,3}$   
	T.~Do,$^{4}$  
 F. Nogueras-Lara,$^{5}$ 
 N. Neumayer,$^{5}$ 
 \newauthor 
C. J. Walcher,$^{6}$ 
A. Seth,$^{7}$ 
R.~Sch{\"o}del,$^{8}$ 
P.~T.~de Zeeuw,$^{9,10}$  
M.~Hilker,$^{11}$ 
Nora L{\"u}tzgendorf,$^{12}$
\newauthor
H.~Kuntschner,$^{11}$ 
M.~Kissler-Patig$^{13}$ \\
$^{1}$The University of Chicago, The Department of Astronomy and Astrophysics, 5640 S. Ellis Ave, Chicago, IL 60637, USA\\
$^{2}$Department of Physics and Astronomy, Michigan State University, East Lansing, MI 48824, USA \\
$^{3}$Department of Computational Mathematics, Science, and Engineering, Michigan State University, East Lansing, MI 48824, USA\\
$^{4}$UCLA Galactic Center Group, Physics and Astronomy Department, UCLA, Los Angeles, CA 90095-1547, USA \\ 
$^{5}$Max Planck Institute for Astronomy, K{\"o}nigstuhl 17, 69117 Heidelberg, Germany  \\ 
$^{6}$Leibniz-Institut f{\"u}r Astrophysik Potsdam (AIP), An der Sternwarte 16, D-14482 Potsdam, Germany\\
$^{7}$Department of Physics and Astronomy, University of Utah, Salt Lake City, UT 84112, USA\\
$^{8}$Instituto de Astrof\'{i}sica de Andaluc\'{i}a (CSIC), Glorieta de la Astronom\'{i}a s/n, 18008 Granada, Spain   \\
$^{9}$Sterrewacht Leiden, Leiden University, Postbus 9513, 2300 RA Leiden, The Netherlands\\
$^{10}$ Max-Planck-Institut    f{\"u}r    Extraterrestrische    Physik,    85748, Garching, Germany\\
$^{11}${European Southern Observatory (ESO), Karl-Schwarzschild-Stra{\ss}e 2, 85748 Garching, Germany}\\
 $^{12}$European Space Agency,
c/o STScI, 3700 San Martin Drive, 21218 Baltimore, MD, USA\\
 $^{13}$European Space Agency - European Space Astronomy Centre, Camino Bajo del Castillo s/n, 28692 Villanueva de la Cañada, Madrid, Spain
}
                                                     
\begin{document}

\date{Accepted 2020 March 5. Received 2020 March 5; in original form 2019 September 18}

\pagerange{\pageref{firstpage}--\pageref{lastpage}} \pubyear{year}

\maketitle

\label{firstpage}

\begin{abstract}
We present stellar metallicity measurements of more than 600 late-type stars in the central 10\,pc of the Galactic centre. Together with our previously published KMOS data, this data set  allows us to investigate, for the first time, spatial variations of the nuclear star cluster's  metallicity distribution.
Using the integral-field spectrograph KMOS (VLT) we observed almost half  of the area enclosed by the nuclear star cluster's effective radius.
We  extract  spectra  at medium spectral resolution, and   apply full spectral fitting utilising the PHOENIX  library of synthetic stellar spectra. 
The stellar  metallicities range from  \mh=--1.25\,dex to  \mh\textgreater +0.3\,dex,  with  most of the stars having super-solar metallicity.
We are able to measure an anisotropy of the stellar metallicity distribution. 
In the Galactic North, the portion of  sub-solar metallicity stars  with \mh\textless0.0\,dex is more than twice as high as in the Galactic South. 
One possible explanation for different fractions of sub-solar metallicity stars in different parts of the cluster is a recent merger event. We propose to test this hypothesis  with high-resolution spectroscopy, and by combining the metallicity information with kinematic data. 
\end{abstract}


\section{Introduction}
The Milky Way nuclear star cluster consists of tens of millions of stars, densely packed in a small  region at the centre of our Galaxy. The cluster extends over  $r_e$= 4-5\,pc \citep{sb,fritz16}, with a mass of about 2$\times10^7$\,M\sun\space \citep{sb,2017MNRAS.466.4040F}. The stars in the nuclear star cluster belong to several stellar populations, they cover ages of few Myr to several Gyr, with most of the stars being  more than 5\,Gyr old \citep{blum03,pfuhl11},  
and  metallicities from sub- to super-solar. Unlike the stellar age, metallicities can be measured directly from  individual stellar spectra. This makes metallicities a useful tool to separate different stellar populations, and understand the formation history of the nuclear star cluster.

The first metallicity measurements of cool stars in the nuclear star cluster consisted of small samples. Due to the high extinction, traditional methods for determining metallicities  with optical spectroscopy  cannot be applied in the Galactic Centre, and for this reason, metallicity measurements are based on  $K$-band or $H$-band spectroscopy.  
\cite{carr00,ramirez00,2009ApJ...694...46D} analysed high-resolution spectra and obtained  solar to slightly super-solar iron-based  [Fe/H] for less than 10 stars, most of them red supergiants.
\cite{cunha07} studied 10 supergiants in the central 30 pc and found a narrow, slightly supersolar metallicity distribution     with enhanced [$\alpha$/Fe] ($\sim$0.2-0.3\,dex). Supergiant stars are relatively young, $\lesssim$1\,Gyr, and rare. 
Red giant stars are better suited to study the metallicity distribution of stars in the nuclear star cluster, as  red giants are older, abundant, and  bright enough for spectroscopic observations. 

In the past years, several studies measured  metallicities of red giant stars in the nuclear star cluster. \cite{dolowfe} analysed spectra of  83 red giant stars with medium spectral resolution  ($R$ $\sim$5,400) and measured the overall metallicity \mh. They found a broad metallicity distribution, ranging from sub-solar  (\mh\textless$-1$\,dex) to metal-rich stars (\mh\textgreater+0.5\,dex). The majority of the stars has super-solar metallicity (\mh\textgreater0\,dex). This finding was confirmed by \cite{kmoslt} on a larger sample of 700 stars and similar methods ($R\sim$4,000). \cite{2017AJ....154..239R} studied a  sample of 17 stars, but higher spectral resolution  ($R\sim$24,000). They measured the iron-based metallicity [Fe/H], and obtained a median value of --0.16\,dex, but also a large spread from sub-solar to super-solar metallicities.
These studies confirm the complex star formation history of the nuclear star cluster.

In order to fully understand the formation and history of the nuclear star cluster, it is important to understand the metallicity distribution of the stars in the cluster, and if there are any spatial variations. We extended our previous work presented in \cite{kmoslt}, where we measured the metallicity distribution of the central 4\,pc$^2$ (radial range 0.1 -- 1.4\,pc) of the nuclear star cluster, by additional data. Both data sets were observed and analysed with an identical observational setup. The new data set presented here covers an area that is larger by a factor 5.6, at  a radial range of 0.4 to 4.9\,pc from the centre.
The combined data sets covers 26.7\,pc$^2$, which is about half of the area enclosed by the effective radius \citep[assuming 4.2\,pc,][]{sb}. The data extend approximately along the Galactic plane, reaching to the effective radius $r_e$ in the North-West and South-East. Our data set  allows us to study, for the first time, spatial  variations of the metallicity distribution within the Milky Way nuclear star cluster. 

This paper is organised as follows: We present our data set in Section~\ref{sec:sec2}, and describe our spectral analysis in Section~\ref{sec:sec3}. In Section~\ref{sec:sec4} we present our results of the metallicity distribution, and discuss them in Section~\ref{sec:sec5}. Our conclusions follow in Section~\ref{sec:sec6}.

\begin{figure*}
  \centering  
  \includegraphics[width=\textwidth]{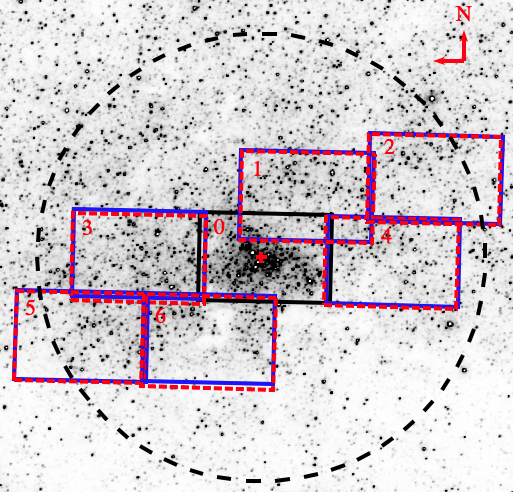}
 \caption{The region covered by our KMOS observations. We observed six fields (labelled from 1 to 6) out to the effective radius $r_e$=110\arcsec=4.2\,pc, which is denoted by the black dashed circle.  The central field 0 was presented  in \citet{kmoset,kmoslt}. 
 The red cross denotes the position of the supermassive black hole.  The underlying image is from the VISTA Variables in the Via Lactea Survey   \citep{2012A&A...537A.107S} in $K_S$ band, Galactic North is up, Galactic East is to the left. The blue and red dashed boxes indicate the repeated observations per mosaic field.}
 \label{fig:observations}
\end{figure*}

\section[]{Data set}
\label{sec:sec2}
\subsection{Observations}
\label{sec:obs}
Our spectroscopic observations were performed with KMOS \citep{kmos} on VLT-UT1 (Antu)  in 2014, in the nights of April 10, 12, 24, May 11, 31, and June 6 in service mode. KMOS is a multi-object spectrograph with 24 integral field units (IFUs), which can be arranged in a close configuration. With 16 dithers in this close configuration, it is possible to observe a mosaic covering 64.9\,arcsec $\times$ 43.3\,arcsec. We observed six mosaic fields of the Milky Way nuclear star cluster, within its half-light radius \citep[$r_e$=110\arcsec=4.2\,pc,][]{sb}.  We chose the location of the fields such that we extended Field 0 \citep{kmoslt}, to obtain approximately symmetric coverage toward Galactic East and West, while avoiding
the region of higher extinction in the Galactic South-West. The region covered by our data is shown in  Fig. \ref{fig:observations}.

Depending on the distance from the centre of the cluster,  we chose exposure times of 155\,s or 190\,s. The exposure time is shorter closer to the centre, to prevent persistence and saturation. 
Each field was observed twice, this means we have in total 12 mosaics. In some of the nights, different IFUs had technical problems and were not used, causing several of the mosaics to have holes. For that reason we observed the two mosaics of the same field with different rotator angles, 120\degr\space and --60\degr\space (except for Field 2, for which both mosaics have the same rotator angle). Rotating the mosaics  by 180\degr\space makes sure that  inactive IFUs fall on different regions of the sky, and we have at least one exposure of each region. We extracted and analysed spectra from each exposure separately.  If we had more than one exposures of a star, we took the mean of the stellar parameter measurements.

We observed in the near-infrared $K$-band ($\sim$19\,340 -- 24\,600\,$\angstrom$), where the spectral resolution is about 4000. The spatial sampling is  0.2\,arcsec\,pixel$^{-1}$\,$\times$\,0.2\,arcsec\,pixel$^{-1}$, the sampling along the dispersion axis is  $\sim$2.8\,$\angstrom$\,pixel$^{-1}$. 
We made offsets to a  dark cloud (G359.94+0.17, $\alpha$\,$\approx$\,266\fdg2, $\delta$\,$\approx$\,$-$28\fdg9,   \citealt{dutra_darkcloud_01}) for sky observations, with the same exposure time as on source. For telluric corrections, we observed B-type dwarf stars. 
We summarise our observations in Table \ref{tab:observations}. We additionally list Field 0, which was observed in September 2013. The observations and results of this field were already presented in \cite{kmoset} and \cite{kmoslt}. This field covers the very centre of the Milky Way nuclear star cluster, and we used it for comparison. 

\setcounter{table}{0}
\begin{table*}
 \centering
  \begin{minipage}{155mm}
\caption{Summary of KMOS observations}
 \label{tab:observations}
\begin{tabular}{@{}cccccccccc@{}}
\noalign{\smallskip}
\hline
\noalign{\smallskip}
Field& Mosaic &Night&Exposure time &Seeing & Rotator angle &Inactive IFU&RA&Dec\\
&	&		& [s] &[arcsec]&[degree]&&[degree]&[degree]\\ \noalign{\smallskip}
\hline
\noalign{\smallskip}
1	& 1&    10. April 2014 & 155&  0.6--0.9   &--60& -- &266.405&	-29.009\\
	& 2&	   10. April 2014  & 155&  0.6--0.9   &120 & - &266.405&	-29.009\\	
2	& 1&    11. May 2014 & 190&  0.9--1.2   &120& 4   & 266.392	&-29.022\\
	& 2&	   11. May 2014  & 190&  0.9--2.5   &120 & 4 &	266.392&	-29.022&\\	
3	& 1&    31. May 2014 & 155&  0.7--1.5   &-60& 4, 11 & 	266.427&	-28.993\\
	& 2&	   6. June 2014  & 155&  0.6--0.8   &120 & 4, 11,15 &266.427&	-28.993	\\
	4	& 1&	   6. June 2014  & 155&  0.7--1.1   &--60 & 4, 11,15 &	266.407&	-29.023\\
		& 2&	   6. June 2014  & 155&  0.9--1.3   &120 & 4, 11,15 &		266.407&	-29.023 \\
5	& 1&    12. April 2014 & 190&  0.9--1.2   &--60& --&   266.442&	-28.992\\
	& 2&	   24. April 2014  & 190&  0.9--1.4   &120 & --&	266.442	&	-28.992& \\
6	& 1&    12. April 2014 & 155&  1.1--1.8   &--60& --  &266.432&	-29.008 \\
	& 2&	   12. April 2014  & 155&  1.0--1.6   &120 & -- &266.432&	-29.008& \\	
 \hline 
0	& 1&    23. Sept. 2013 & 100&  0.7--0.9   &120& 13 & 266.417&-29.007\\
	& 2&	   23. Sept. 2013  & 100& 1.0--1.4   &120 & 13 &266.417&-29.007\\	
 \hline  \end{tabular}
\end{minipage}
\end{table*}

\subsection{Data reduction}
\label{sec:reduction}
Our data reduction procedure is similar to the reduction of the central field, described in \cite{kmoset}. 
We used the KMOS pipeline \citep{spark} provided by ESO with EsoRex (ESO Recipe Execution Tool). The reduction steps include dark subtraction, flat fielding, wavelength calibration, illumination correction  using the flat field exposures, and telluric correction with a standard star. Before telluric correction, we removed the intrinsic stellar absorption lines and the blackbody spectrum from the  standard star spectrum with our own  \textsc{idl} routine. After these reduction steps, the object and sky exposures were reconstructed  to data cubes. We combined the two sky frame exposures of each observing block to a mastersky frame, and   subtracted  the sky by scaling it to the object cubes, as described by \cite{ohsky}. Cosmic rays were removed  with an  \textsc{idl} program for data cubes  \citep{spark} based  on the program \textsc{l.a.cosmic} \citep{lacosmic}.

We have complimentary photometric catalogues in $J$, $H$, and $K_S$-bands in the field, and  extinction maps provided by \cite{exthawki,2019A&A...631A..20N}. The imaging observations were done with HAWK-I in speckle holography mode, which ensures a high spatial resolution and completeness. For 18 bright stars with no HAWK-I photometry due to saturation, we used SIRIUS $K_S$ band photometry \citep{shogo06}.  The star catalogue is used to extract stellar spectra. For the central field \citep[see][]{kmoset,kmoslt}, we used the program \textsc{pampelmuse} \citep{pampel}, however,  for the observations analysed in this paper, this was not feasible.  \textsc{pampelmuse} performs PSF fitting and light deconvolution in crowded fields.  It requires exposures of several stars to fit the PSF.  Due to the lower  stellar density compared to the central field,   there were   not enough stars  in a single exposure for a PSF fit. We could not combine several subsequently taken exposures, as observing conditions varied  too much. 
Instead of using \textsc{pampelmuse}, we only extracted stars that are isolated from nearby stars to avoid blending of extracted spectra. We used a $k$-d-tree algorithm to identify stars in the photometric catalogue
 with no neighbour within   six KMOS pixels (i.e. 1\farcs2), unless the neighbour stars are fainter by at least 3\,mag in   the $K_S$ band. We extracted spectra of such stars within a circular aperture with a 3\,pixel (0.6\arcsec) radius by simply adding the flux within the aperture. We also subtracted the background flux determined from a 2\,pixel (0.4\arcsec) wide annulus, at a radius \textgreater 4\,pixel (0.8\arcsec).  For a fair comparison,  we used this  method to re-extract spectra in the central field and re-analyse the spectra. We found consistent results with \cite{kmoslt}.  
Depending on the stellar density of the field and the number of active IFUs at the time of the observations, we extracted about 190-320 spectra per individual mosaic.
We note that these numbers include foreground stars,  multiple exposures of the same star, and  low signal-to-noise spectra.  We corrected the velocity scale of the spectra to the local standard of rest.

As noted by \cite{2015ApJ...805..182G}, the spectral resolution of the 24 KMOS IFUs varies spatially, both for different IFUs and for  individual IFUs, with a mean value of $R$ = $\lambda / \Delta \lambda$ = 4\,200. Within one IFU the spectral resolution $R$   can have a standard deviation of up to 150, and over all IFUs the standard deviation is about 300.  We measured the line-spread function on the reconstructed sky data cubes, as described in \cite{kmoslt}. We fitted Gaussian functions to several sky lines in the wavelength region $\lambda$\,=\,21\,900 --\,22\,400\,$\angstrom$, and created spatially resolved resolution maps for the 24 IFUs. These will be used for the full spectral fitting performed in Sect. \ref{sec:fitting}. 

\section{Analysis}
\label{sec:sec3}
In this section, we describe the different analysis steps to measure stellar parameters. We derived  radial velocities using a large spectral range  before we measured spectral indices. Then, we applied full spectral fitting to derive the stellar parameters metallicity \mh, effective temperature $\teff$, and surface gravity $\logg$. 

 \subsection{Measuring kinematics and spectral indices}

We fitted the stellar spectra in the wavelength  region 20\,880 to 23\,650\,$\angstrom$ with the  \textsc{idl} program \textsc{pPXF} \citep{ppxf}  to measure the stellar radial velocity. 
We used  the high resolution spectra of late-type stars by \cite{wallace} as templates, and convolved them to the mean spectral resolution of the KMOS spectra.  The stellar spectra contain several gas emission lines, which originate from the interstellar gas inside the Milky Way nuclear star cluster: 
In the central 2\,pc of the Milky Way nuclear star cluster, there is the so-called ``minispiral'' or Sgr~A West \citep[see e.g.][]{paumard04,2012A&A...538A.127K}. It is visible in  the   \mbox{Br\,$\gamma$} (21\,661\,$\angstrom$) and \mbox{He\,{\sc i}} (20\,587\,$\angstrom$) transitions in emission (see Figs. 7 and 8 in \cite{kmoset}). Further out, in the central 6\,pc of the Milky Way, there is a clumpy circum-nuclear ring \citep[see e.g.][]{2012A&A...542L..21R,isaacanja}, which emits at several \mbox{H$_2$} transitions, e.g. 21\,218\,$\angstrom$, 22\,235\,$\angstrom$, 22\,477\,$\angstrom$. We masked the wavelength regions of emission lines and  several sky emission lines in the \textsc{ppxf} fit. 
As result we obtained the radial velocity of the star, and additionally the velocity dispersion. The velocity dispersion has no physical meaning.  As we fit single stars, it is usually low. However, it can reach high values, which indicates a bad fit of a low signal-to-noise spectrum. We measured the uncertainties by running Monte Carlo simulations, and adding noise to the spectra. The radial velocity was later used as prior information for the full spectral fitting.

The strength of different absorption lines can be used to estimate effective temperatures, and thus differentiate cool late-type stars from hot early-type stars. Further, it is possible to differentiate   red supergiant from red giant stars. We measured spectral indices to constrain the possible ranges of  effective temperature and surface gravity.  We measured the equivalent width of the first CO band head ($\sim$22\,935~$\angstrom$), and  the \mbox{Na\,{\sc i}} doublet (22\,062~$\angstrom$ and 22\,090~$\angstrom$) 
with the index definitions of \cite{frogel}, after correcting the spectra to the  rest frame.

\subsection{Full spectral fitting}
\label{sec:fitting}
\begin{figure*}
  \includegraphics[width=0.95\textwidth]{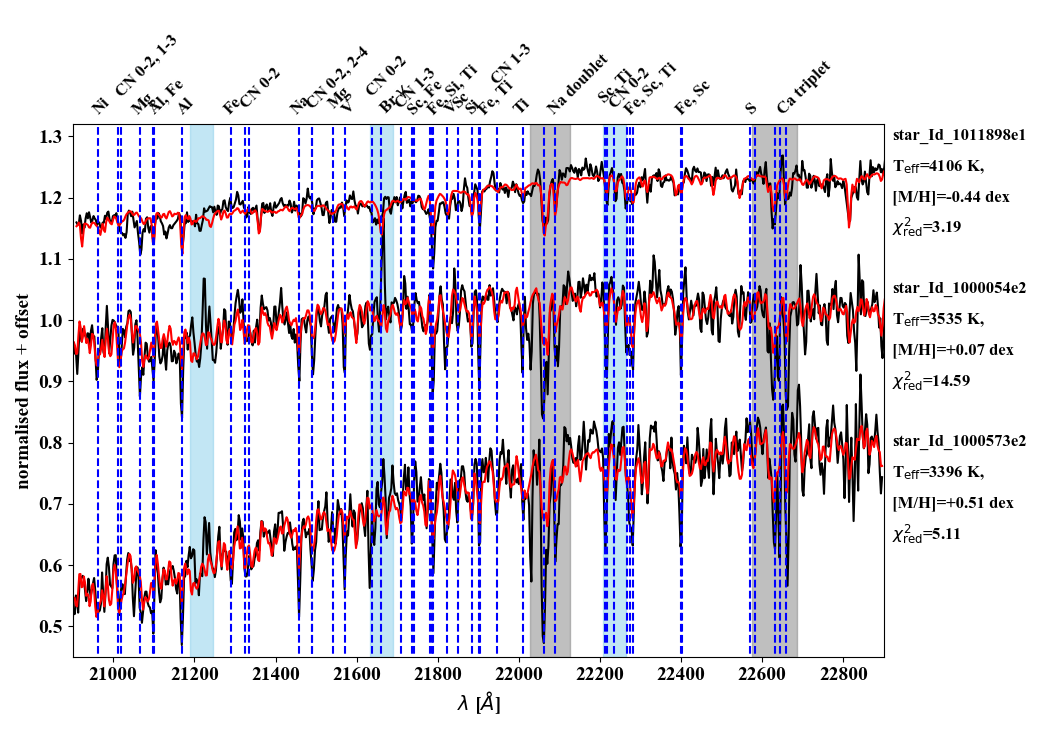}
\caption{Spectra (black) and best-fitting model (red) of three different stars with metallicities \mh=-0.44, +0.07, and +0.51\,dex. Blue dashed lines denote known spectral lines, labelled on top of the plot. Grey shaded areas were excluded from the fit, the light blue shaded areas were excluded for spectra with strong contamination of the Brackett-$\gamma$ or  H$_2$ gas  emission lines. }
\label{fig:spec}
\end{figure*}
In order to constrain the stellar parameters, we fitted the stellar  spectra of our KMOS data set. We used  the \textsc{StarKit} code developed by  \cite{starkit}. This code was also used by \cite{dolowfe} and \cite{kmoslt}. 
\textsc{StarKit}  applies  Bayesian sampling  \citep{multinest,multinestpy}. 
The code uses a grid of synthetic spectra and interpolates them, to   find the best-fitting stellar parameter to a stellar spectrum.  

As in \cite{kmoslt}, we used the PHOENIX spectral library \citep{husserphoenix} of synthetic spectra. The synthetic spectra   are in a grid with  $\teff$\,=\,[2\,300 K; 12\,000 K] and a step size of $\Delta \teff$\,=\,100\,K, \mh\,=\,[$-$1.5\,dex, +1.0\,dex], $\Delta$ \mh\,=\,0.5\,dex, $\logg$\,=\,[0.0\,dex, 6.0\,dex],  and $\Delta \logg$\,=\,0.5\,dex. \mh\space denotes the overall metallicity of all elements, not the Fe-based metallicity. The model spectra have  [$\alpha$/Fe]=0, but [$\alpha$/H]=\mh. We were not able to measure [$\alpha$/Fe] as additional fitting parameter.  Our tests resulted in sub-solar to solar  [$\alpha$/Fe], also for stars with sub-solar [M/H]. We do not consider these results reliable, and believe that our spectral resolution is too low. Most absorption lines are blends of several elements, which makes it hard to constrain single element abundances. Therefore, we decided to fit only \mh, \teff, and \logg, and assume [$\alpha$/Fe]=0. The effect of nonzero  [$\alpha$/Fe] on our measurements is included in our systematic uncertainties. 

The model spectra were convolved to the respective spectral resolution of each KMOS spectrum, as determined by the sky lines on the location on the respective IFU (see Sect. \ref{sec:reduction}). 
We fitted the effective temperature  $\teff$, metallicity  \mh, surface gravity $\logg$, and radial velocity $v_z$. The fits were done in the wavelength region  $\lambda$\,=\,20\,900 --\,22\,900\,$\angstrom$. 
Some stellar spectra also have gas emission lines at  \mbox{Br\,$\gamma$} (21\,661\,$\angstrom$) and \mbox{H$_2$} transitions at 21\,218\,$\angstrom$, and 22\,235\,$\angstrom$. We excluded the region around  the emission lines  from the fit if the emission line region had a significantly higher standard deviation  than the  rest of the spectrum. 
Further, we excluded the regions of the spectrum where the   Na and Ca lines are ($\lambda$\,=\,[22\,027\,$\angstrom$, 22\,125\,$\angstrom$] and $\lambda$\,=\,[22\,575\,$\angstrom$, 22\,685\,$\angstrom$]), as  Galactic centre stars have stronger Na and Ca lines  compared to normal disc stars \citep{blum96,kmoslt}, which biases the fit to unrealistic high metallicities. We show three spectra and their best-fit models as examples in Fig.~\ref{fig:spec}.  
The spectral continuum shape is influenced by extinction, which can bias the fit of the stellar parameters. For that reason the continuum was  modelled with a fifth degree polynomial function.

We used the radial velocity measured with \textsc{ppxf} as prior information, with a Gaussian prior. The \textsc{ppxf} radial velocity was set as mean of the Gaussian, and the radial velocity uncertainty as the width of the Gaussian. 
The magnitude of a star contains information about its luminosity class, and thus constrains the possible range of the surface gravity. We corrected the magnitudes for extinction using the extinction map of \citet[][Fig. 30]{exthawki}. We used the extinction corrected $K_{S,0}$-band magnitude to set constraints on the surface gravity, as done by \cite{dolowfe} and \cite{kmoslt}:
Since brighter stars have a lower surface gravity, we set the uniform priors for stars with   $K_{S,0}$\,\textless\,12~mag to 0.0\,dex\,\textless\,$\logg$\,\textless 4.0\,dex, and for stars with $K_{S,0}$\,$\geq$\,12\,mag to 2.0\,dex\,\textless\,$\logg$\,\textless\,4.5\,dex. A further constraint comes from the equivalent width of the CO absorption line. Stars with $EW_{CO}$\,\textgreater\,25\,$\angstrom$ are potentially supergiants. We set the prior uniform to  0.0\,dex\,\textless\,$\logg$ \textless\,2.0\,dex for stars with $EW_{CO}$\,\textgreater\,25\,$\angstrom$ and $K_S$\,$\leq$\,10\,mag, and to 0.0\,dex\,\textless\,$\logg$ \textless\,4.0\,dex for stars with $EW_{CO}$\,\textgreater\,25\,$\angstrom$ and $K_S$\,\textgreater\,10\,mag. 

The effective temperature and metallicity priors  were set uniform in the ranges  [2300\,K, 12000\,K] and  [$-$1.5\,dex, +1.0\,dex], respectively.  
We also tested a Gaussian prior for $\teff$ on a subset of stars, by using $\teff$ determined from the empirical  $EW_{CO}$--$\teff$ calibration derived in \cite{kmoslt}. This increased the fitting results of $\teff$ by a median value of 33\,K, whereas the median changes of the other measurements, $\logg, \mh$ and $v_z$ were close to zero. For easier comparison with the results of \cite{kmoslt}, where a uniform prior for $\teff$ was used, we decided to use the uniform prior in this study as well.  But this test shows that the metallicity results are robust under moderate $\teff$ variations.

\subsection{Data selection}
\label{sec:selection}

To measure the stellar parameters, we require high signal-to-noise. We  excluded stars with low signal-to-residual-noise (\textless$\sncut$) spectra,  or large fitting uncertainties  ($\sigma_{ \teff}$\textgreater 250\,K, $\sigma_{\mh}$\textgreater 0.25\,dex, $\sigma_{\logg}$\textgreater 1\,dex, $\sigma_{v_z}$\textgreater 10\,km\,s$^{-1}$). 
We combined the fit results obtained from  individual spectra of the same star  to a mean stellar parameter measurement.  
We have \goodstar\space  stars with at least one good stellar parameter fit, from 1136 analysed spectra.

\begin{figure}
    \centering  
  \includegraphics[width=\columnwidth]{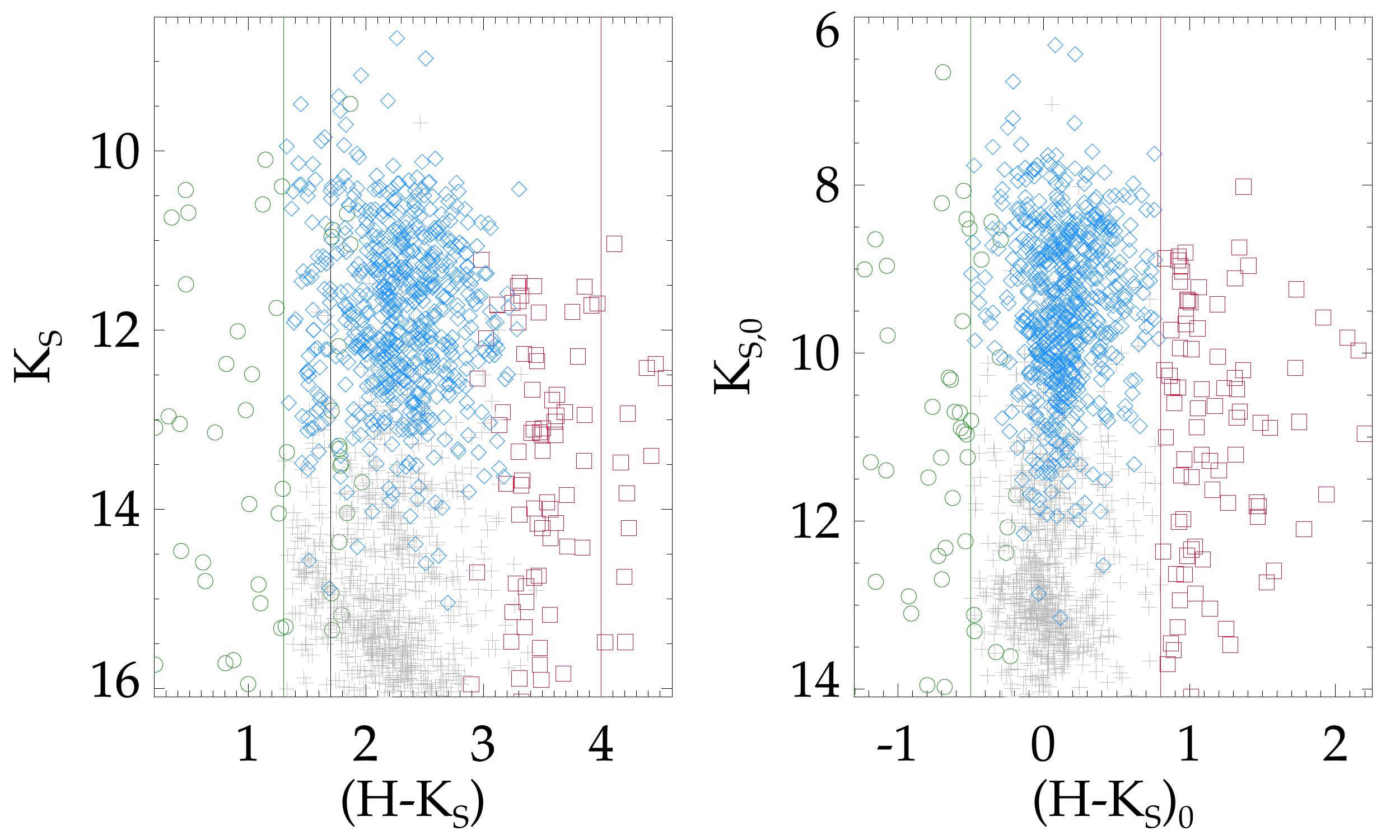}
 \caption{Colour-magnitude diagrams of  stars within the KMOS Fields 1-6 with extracted spectra and $H$ and $K_S$ photometry, before (left) and after (right) extinction correction. Our sample of member stars is shown as blue diamond symbols. We classify stars with observed colours $H-K_S$\textless 1.3\,mag or extinction corrected colours $(H-K_S)_0$\textless--0.5\,mag as
 foreground stars (green circles), stars with  $H-K_S$\textgreater 4.0\,mag or  $(H-K_S)_0$\textgreater 0.8\,mag as background stars (red squares). Grey plus signs denote stars that were excluded due to our quality cuts on signal-to-noise and statistical uncertainties,  red and green vertical lines show our colour cuts for member stars. We applied a different extinction map on stars left of the black vertical line at 1.3\,mag\textless$(H-K_S)$\textless1.7\,mag (see text for details).}
 \label{fig:cmd}
\end{figure}

However, this set includes also foreground stars. 
We used the photometry to determine which stars are members of the Milky Way nuclear star cluster. In particular, the  extinction corrected $(H-K_S)_0$ colour allows  to identify foreground stars. The intrinsic $H-K_S$ colour of late-type stars in the nuclear star cluster is in the rather narrow range of about  [$-$0.13\,mag, +0.38\,mag] \citep{do13,rainer_review14}. This holds for  stars in our magnitude range,  for metallicties  from -1.0 to 0.6\,dex, and for ages from 6.5 to at least 10\,Gyr \citep{bressan12,chen14,tangiso14,2016ApJ...822...73R}. 
If a star has a significantly bluer extinction-corrected colour $(H-K_S)_0$, it is an over-corrected foreground star. 
We consider stars with extinction-corrected $(H-K_S)_0$\,\textless\,$-$0.5\,mag  as foreground stars. To correct for extinction, we used the extinction maps of \cite{exthawki}, in particular, we corrected stars with observed $H-K_S$\textless 1.7\,mag with the extinction map derived from  stars with 1.4\,mag\,\textless$H-K_S$\textless 1.7\,mag, and stars with observed $H-K_S$\textgreater 1.7\,mag with the extinction map derived from  stars with 1.7\,mag\,\textless$H-K_S$\textless 3.0\,mag, as shown by \cite{exthawki}. 
In addition, we consider stars with uncorrected $H-K_S$\textless 1.3\,mag as foreground stars, following \cite{2019NatAs.tmp....4N}.
On the other hand, stars with a redder colour are probably subject to a higher extinction, and are thus potential background stars. We classify a star as potential background star if its extinction corrected colour $(H-K_S)_0$ \textgreater \bgcut\,mag. 
We show a colour-magnitude diagram of our data set in Fig.~\ref{fig:cmd}.  

For  31 stars  we have only one band, $H$ or $K_S$, or they lie in a region of the extinction map where extinction is underestimated \citep[b\textless-0.06 degree,][]{exthawki}. Hence we cannot classify them as member stars using photometry. However, foreground stars have a steeply decreasing continuum in the $K$-band spectrum, whereas member stars have a rather straight or even increasing continuum slope. Thus, we can infer the member status of a given star from the spectrum. We measured the continuum slope with  an outlier-resistant two-variable linear regression  fit in the wavelength region $\sim$19\,600--22\,000\,$\angstrom$. Then, we applied machine learning algorithms to the data set. The training set consists of about 800 stars with known membership status from using the  photometry and extinction map. We included central stars (Field 0) to the training set. We  used the \textsc{R entropy} package to select useful machine learning variables. The by far most important variable is the spectral continuum slope, however, also the $y$-intercept, the Na and CO equivalent widths, RA and Dec have a small effect on the machine learning result.  This can be expected, as these properties are different for Galactic center stars than for foreground stars. In particular,  Galactic centre stars have rather high $EW_{Na}$  \citep{blum96,kmoslt} compared to normal disk stars. At the Galactocentric distance of 8\,kpc, the member stars of our sample are M giant or supergiant stars rather than main sequence stars. This means that $EW_{CO}$  is  larger in member stars compared to foreground stars, which  can be earlier giant stars or main sequence stars. The position of the stars on the sky has a small effect on the outcome of the machine learning, the radial velocity has the highest entropy of all considered parameters (larger by a factor 4). This means that the radial velocity cannot  differentiate foreground stars from member stars and we do not use it as variable.

We use 10-fold cross-validation and average the results to determine the  classifier.  We tested various classification methods with the \textsc{r} package, and found 
that fitting Multinomial Log-linear Models via neural networks (function \textsc{multinom} in package \textsc{nnet}) 
 has the smallest misclassification error, 6.4\%. 
Our focus is a  small false positive rate (FPR), which denotes the ratio of foreground stars that are misclassified as member stars divided by the total number of foreground stars,  rather than  the  false negative rate (FNR), which  is the ratio of member stars that are misclassified as foreground stars divided by the total number of member stars, because we rather discard a member star than including a foreground star in our sample. The \textsc{multinom} classifier has FPR=17.6\% and FNR=6.2\%. 
 We apply the \textsc{multinom} classifier on the 31 stars for which we cannot determine the membership status from the photometry. We add the 29 stars  which are classified as member stars to the data set.  With our FPR of 17.6\%, it is unlikely that all the stars classified as member stars are misclassified foreground stars. We expect at most five misclassified foreground stars among the 29 stars classified as member stars. These stars are exclusively located in the region b\textless-0.06 degree, where we have no coverage from the extinction map.  
 
There may still be contamination by stars of  other Galactic components, such as the bar or halo interlopers, that are close to the nuclear star cluster and have a similar colour. We estimate the remaining contamination rate  with  the Besan\c{c}on Galaxy Model\footnote{\url{https://model.obs-besancon.fr/modele_home.php}} \citep[BGM,][]{2014A&A...564A.102C}, assuming a diffuse extinction  of 3.5\,mag$\cdot$kpc$^{-1}$, in  an area of  1\,deg$^2$ around the nuclear star cluster. We consider   the magnitude range 9.48\,mag\textless K\textless 13.19\,mag, which covers the observed $K$-band magnitudes of stars considered in  Sec.~\ref{sec:aniso} and Fig.~\ref{fig:metfracvor}. With these parameters, the BGM contains only one star at a distance of 83 pc from  Sgr~A*, 40 \,pc behind it along the line-of-sight, which  belongs to the young thick disk. All other 469 stars of the BGM are \textgreater638\,pc distant from Sgr~A* along the line-of-sight. For this reason, these stars have a different colour and can be identified as foreground stars. Considering our field-of-view of 4.9\,arcmin$^2$, the model  predicts only 0.00136 stars that may be misidentified as member stars of the nuclear star cluster. The BGM therefore suggests that the remaining  foreground star contamination in our data set is negligible.

\subsection{Uncertainties}
\label{sec:unc}

The full-spectral fitting gives statistical uncertainties $\sigma_\text{fit}$ for the stellar parameters. However, these can be lower than the standard deviation $\sigma_\text{sd}$ from fitting several spectra of the same star. If this was the case, we used the standard deviation $\sigma_\text{sd}$ of the $\sim$360 stars with several exposures rather than the formal fitting uncertainties $\sigma_\text{fit}$. 
For the remaining $\sim$350 stars with only one exposure, we used the median of $\sigma_\text{sd}$ as statistical uncertainty, if it was larger than the formal fitting uncertainty $\sigma_\text{fit}$.

In addition, we considered systematic uncertainties.  \cite{kmoslt} fitted spectra from different stellar libraries with reference stellar parameters using \textsc{starkit}. They found that the \textsc{starkit} results differ by  $\langle \Delta\,\teff \rangle$\,=\,$-$58\,K, $\langle \Delta\,{\mh}  \rangle$\,=\,$-$0.1\,dex, and $\langle \Delta\,{\logg} \rangle$\,=\,0.2\,dex from the reference stellar parameters,  with  standard deviations $\sigma_{\Delta \teff}$\,=\,205\,K,    $\sigma_{\Delta \mh}$\,=\,0.24\,dex, and $\sigma_{\Delta \logg}$\,=\,1.0\,dex. 
The   offsets and scatter  are partially caused by systematics in the model spectra,  by the different alpha-abundances of the library stars,  and  by the different methods and assumptions that were made to derive the  reference stellar parameters. 
Nevertheless, we use the standard deviations as systematic uncertainties and added them in quadrature to the statistical uncertainties. 
We note that the systematic uncertainties were derived by fitting stars with \mh\,\textless\,0.3\,dex. The uncertainties for the stars with higher metallicities may be underestimated. 
The mean total uncertainties are $\sigma_{\teff}$\,=\,212\,K,  $\sigma_{\mh}$\,=\,0.26\,dex, and $\sigma_{\logg}$\,=\,1.0\,dex.
 
 As additional test of systematic uncertainties, we fitted six red giant star spectra of NGC 6388 observed with SINFONI \citep{2013ApJ...769..107L} at a similar spectral resolution as our data. We obtained $\langle \mh \rangle$\,=\,$-$0.54\,dex, which is in agreement with other measurements, and the value listed in the Galactic Globular cluster catalog by \cite{harris,2010arXiv1012.3224H} of [Fe/H]=$-$0.55\,dex.  The six metallicity measurements have   a standard deviation $\sigma_{\mh}$\,=\,0.15\,dex. This value means that,  for a monometallic stellar population, our method will have a dispersion of $\sim$0.15\,dex, which is less  than our systematic uncertainty.  

 \cite{2017AJ....154..239R}  observed 17 M giants at high spectral resolution in the Galactic centre. We matched our data set to theirs and found three  stars, which are probably the same: Their GC13282, GC11025, and GC16887 correspond to our stars Id134, Id1011914, and Id3021083. The samples have  only small differences of the observed velocities (2--8.9\,km\,s$^{-1}$),  $K_S$  (0.06--0.29\,mag), and the coordinates (0.27--0.78 arcsec). Our results for \teff\space and \logg\space agree well within the uncertainties, though our results for \logg\space are lower in all three cases. The metallicity is harder to compare, as we measured the total metallicity \mh, while \cite{2017AJ....154..239R} measured the iron-based metallicity [Fe/H].  But if we ignore this, our results  agree within the uncertainties.  Furthermore, we observed the same trend, with increasing metallicity from Id134 (\mh=0.19\,dex) over Id1011914 (0.23\,dex)  to Id3021083 (0.56\,dex).   
 
\subsection{Completeness}
\label{sec:comp}
The observations were taken at different nights,   at different conditions and exposure times. Thus, we expect that the different fields have a different depth. Also, the foreground extinction varies over the different fields, meaning that we can observe deeper into the Galactic centre, and reach intrinsic fainter stars in regions with less extinction. Another factor is  crowding, we did not extract spectra of stars that had a close neighbour in order  to obtain a clean aperture extraction, and this concerns less stars in the outer part of the cluster. These factors have to be considered when comparing the stellar populations in different regions. 

In order to estimate the completeness, we compared the cumulative distribution of observed $K_S$ of our sample with the $K_S$ distribution of the photometric catalogue. Our sample contains only stars for which we could extract a spectrum with sufficiently high  signal-to-noise-ratio to measure stellar parameters (see Sect.~\ref{sec:selection}), which leaves only $\sim$740 stars, including foreground stars. The photometric catalogue can be considered complete compared to our spectroscopic sample, as it is almost 100\% complete at $K_S$=15\,mag beyond the central parsec \citep{exthawki}. For the different KMOS mosaic regions, we compared the photometric distributions and determined the $K_S$-magnitude at which our sample is 50\% complete. The resulting magnitudes are listed in Table \ref{tab:sample}. We determined the uncertainty of 0.1\,mag by trying different bin sizes of the photometric histograms, and using the standard deviation as uncertainty. The completeness may not be constant over a given field,  as some regions were observed only once due to inactive IFUs, other regions were observed twice. Our completeness limits are therefore averages for a given field. Most fields reach the 50\% completeness limit at 11.9--12.3\,mag. Only Field 2 has a higher completeness, reaching 12.5\,mag. We also  consider the mean extinction $\left<A_{K_S}\right>$ of our stars in  each field, and correct the $K_S$ completeness limit.   This is not a measure of an extinction corrected completeness, but  allows a rough estimate of the completeness variation in the different fields.
With the mean extinction  correction, the completeness ranges from 9.53-10.25\,mag for Fields 1-2, while Fields 3-6 are rather comparable, with 9.71-9.94\,mag. 

We note that Field 0 reaches the 50\% completeness limit at 13.7\,mag (measured beyond the extremely dense central $r$\textless 0.5\,pc), despite the shorter exposure time. This is caused by the different extraction method, which allows to extract faint stars nearby bright stars. However, this method was not feasible for the Fields 1-6 (see Section \ref{sec:reduction}). 
\begin{table}
 \centering
 \caption{Completeness  of analysed sample}
 \label{tab:sample}
\begin{tabular}{@{}ccccc@{}}
\noalign{\smallskip}
\hline
\noalign{\smallskip}
Field& $K_S$50\% complete& $K_S$50\% complete-$\left<A_{K_S}\right>$& $\left<K_{S}\right>$& $\left<A_{K_S}\right>$\\
&	[mag]&[mag]&[mag]&[mag]\\
 \noalign{\smallskip}
\hline
\noalign{\smallskip}
1	&11.9$\pm$0.1 &9.53 &11.7&2.38\\
2	&12.5$\pm$0.1 &10.25 &12.2&2.27\\	
3	&11.9$\pm$0.1 &9.67 &11.8&2.23 \\
4	&12.3$\pm$0.1 &9.87 &12.0& 2.44 \\
5	&12.2$\pm$0.1 &9.94 &12.0&2.27\\
6	&12.2$\pm$0.1 &9.71 &11.7&2.44\\
\hline
0	&13.7$\pm$0.1 &11.42 & 13.1&2.28\\	
 \hline  \end{tabular}
\end{table}


\section{Results}
\label{sec:sec4}

We measured the overall metallicity \mh\space of a sample of \goodstarmemberadd\space stars in the Milky Way's nuclear star cluster, at  projected radii of  0.4 to 4.9\,pc from the central supermassive black hole. 
The stars are located towards the Galactic North-West and South-East, and spread out over an area of \textgreater 22\,pc$^2$. 
We will publish a table of our stellar parameter measurements online.

\begin{figure}
  \centering  
  \includegraphics[width=0.99\columnwidth]{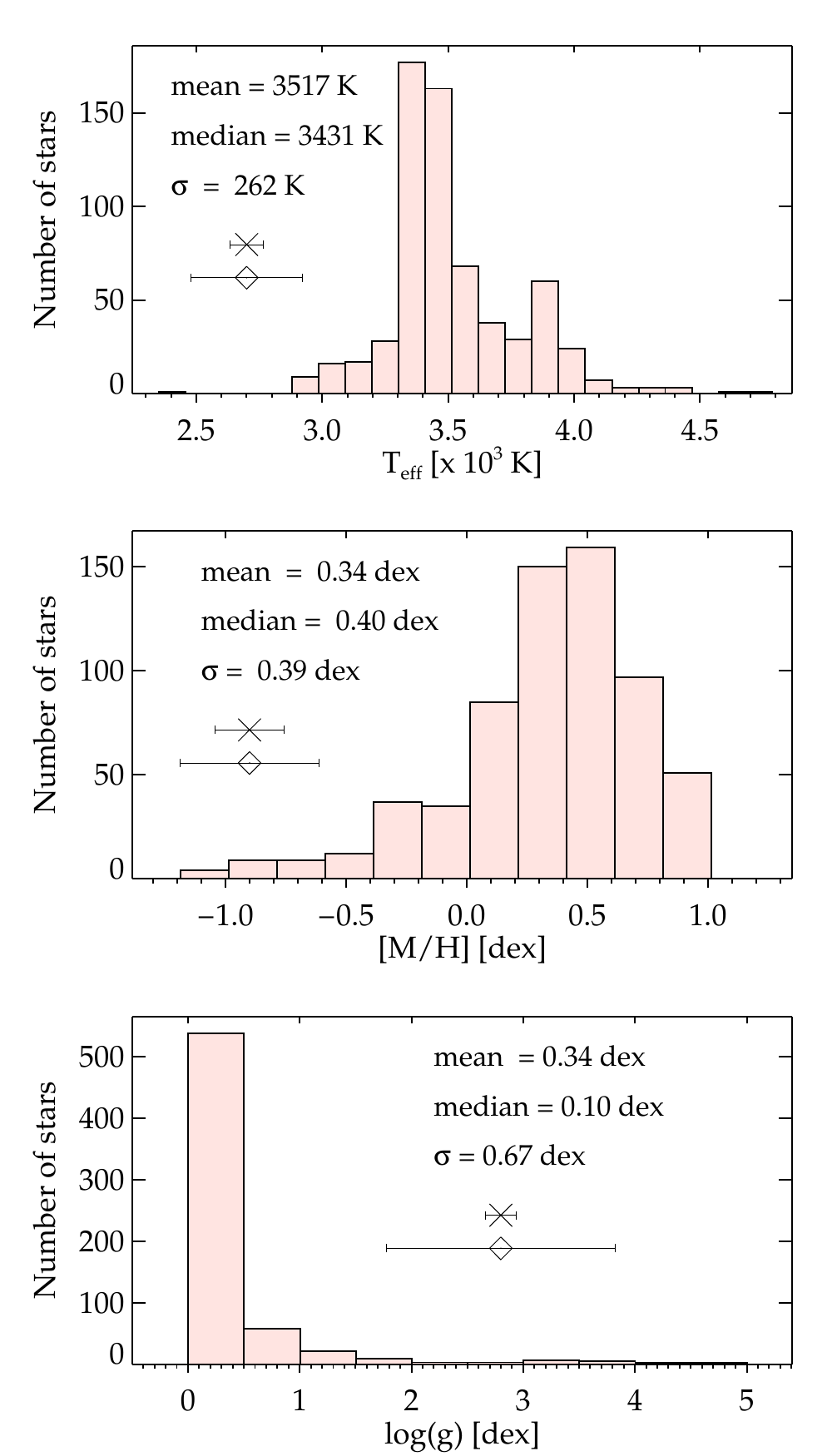}
   \caption{Histograms of the stellar parameters for \goodstarmemberadd\space stars with intrinsic colour within $(H-K_S)_0$\,=\,[$-$0.5\,mag; 0.8\,mag] and observed colour $H-K_S$\textgreater 1.3\,mag. From top to bottom: effective temperature $\teff$,  metallicity \mh, surface gravity $\logg$. Mean, median and standard deviation of the distributions are noted as legend in the respective panels. The error bars denote the mean statistical  uncertainty (cross symbol) and the mean total uncertainty (diamond symbol) of the measurements.}
 \label{fig:meanhist}
\end{figure}

\begin{figure}
 
  \includegraphics[width=0.98\columnwidth]{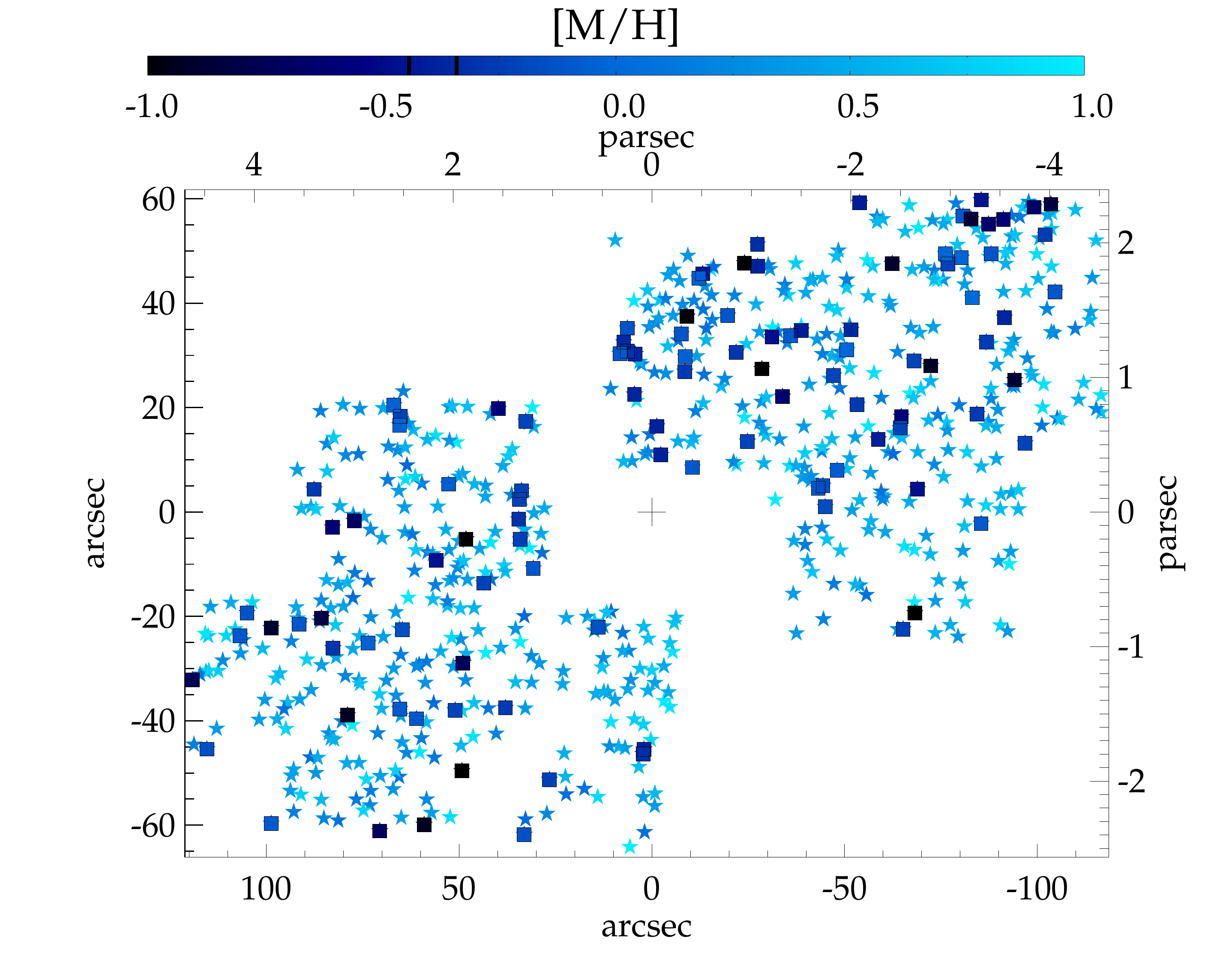}
  \includegraphics[width=0.98\columnwidth]{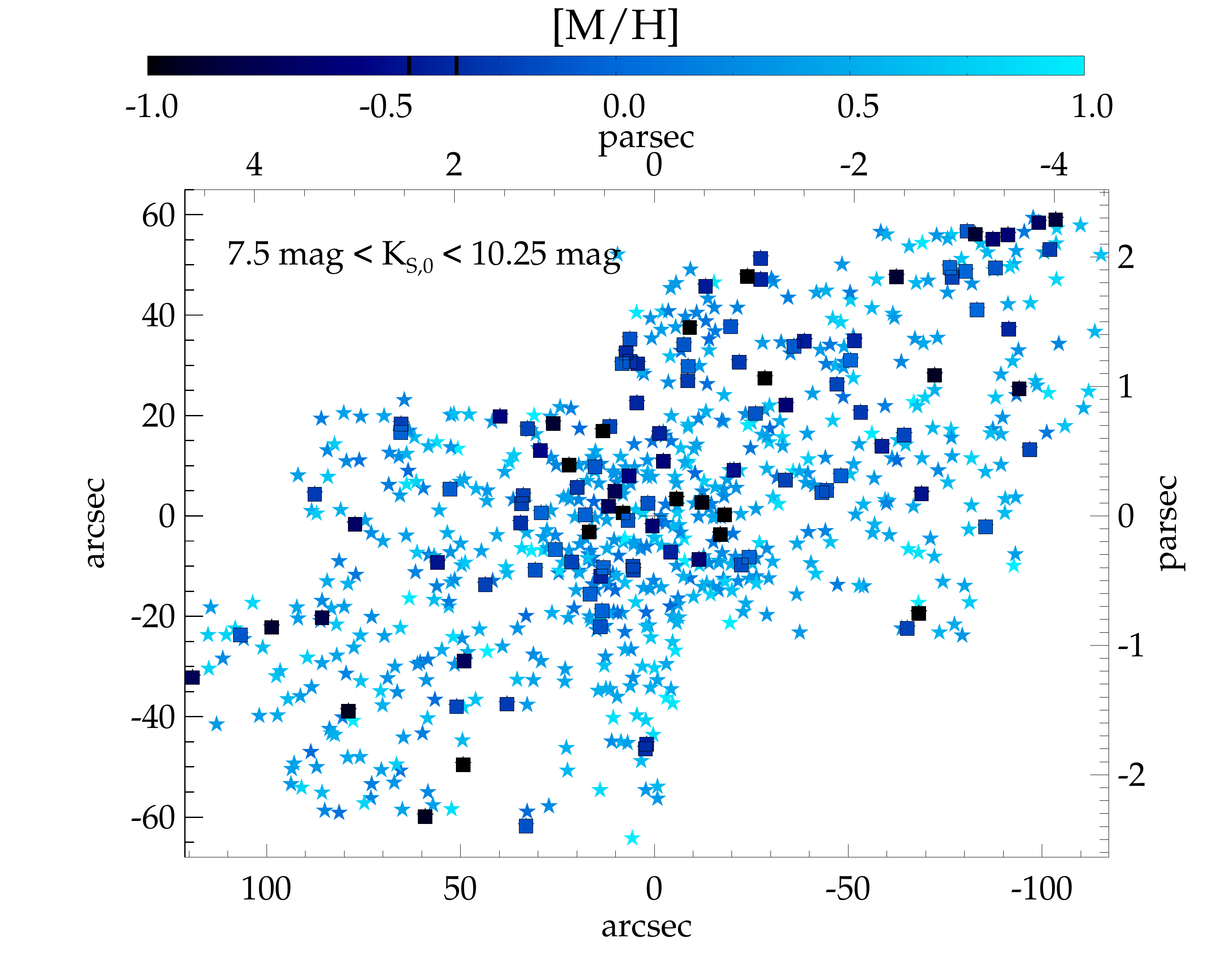}
 \includegraphics[width=0.98\columnwidth]{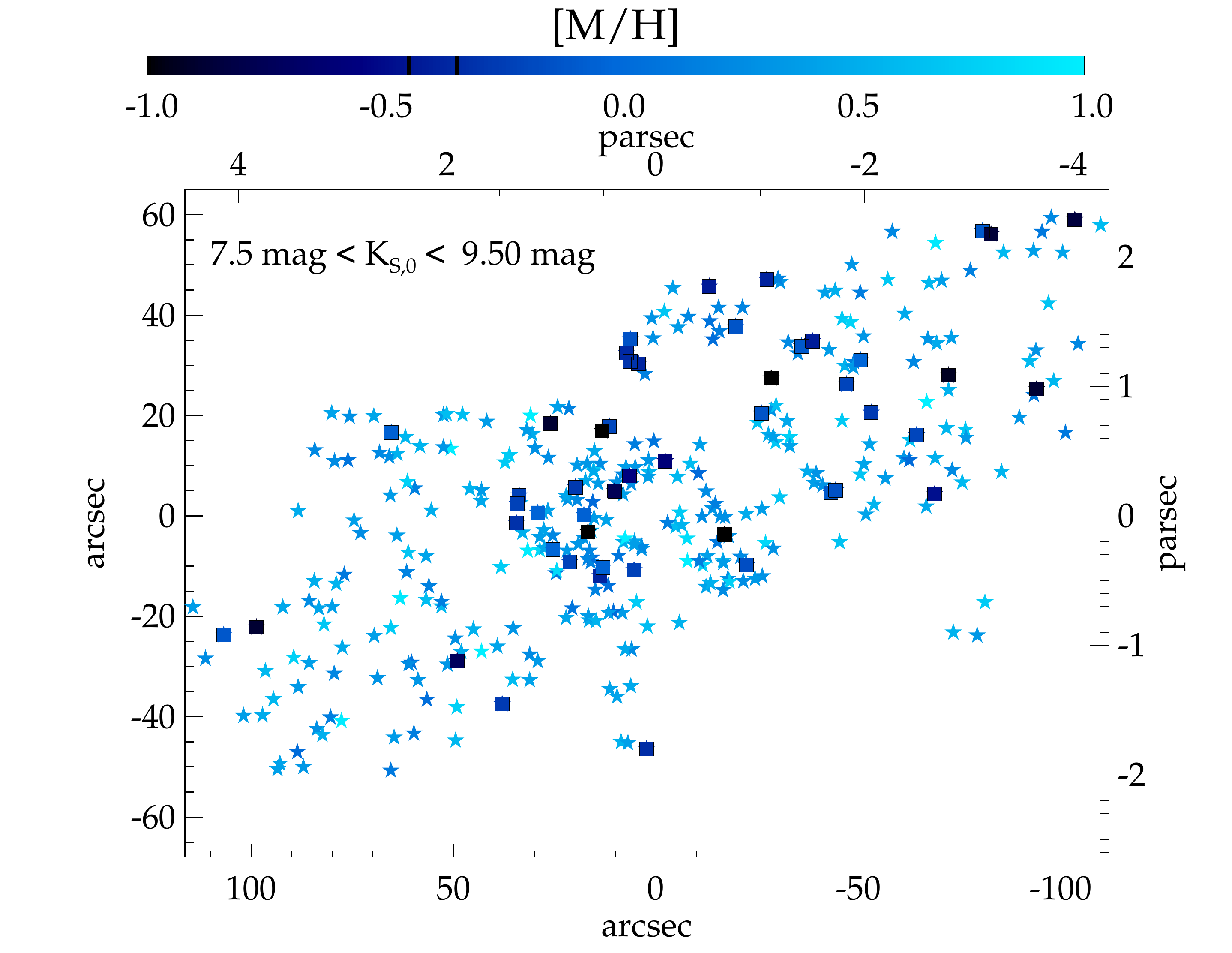}
 \caption{Spatially resolved distribution of the  metallicity   in offset coordinates from Sgr~A* (marked as a black plus sign), Galactic North is up. The different colours denote the metallicity \mh. Upper panel:  \goodstarmemberadd\space stars of our new data set with $(H-K_S)_0$\,=\,[$-$0.5\,mag; 0.8\,mag] and $H-K_S$\textgreater 1.3\,mag; Middle panel:  including the data of Feldmeier-Krause et al. (2017), but using only the 729 stars with extinction corrected $K$-band magnitudes $K_{S,0}$ in the range 7.5--10.25 mag; Lower panel: 344 stars with extinction corrected $K$-band magnitudes in the range 7.5--9.5 mag, and after applying cuts as described in the text. 
  We highlight  sub-solar metallicity stars with \mh\textless 0.0\,dex with  square symbols, they are predominantly located in the centre and towards the North. } \label{fig:metfield2}
\end{figure}


\subsection{Stellar parameter distributions}

We show the stellar parameter distribution   for \goodstarmemberadd\space stars, which are likely cluster members based on their photometry or spectral slope, in Fig.~\ref{fig:meanhist}.  The mean,  median and standard deviation values of the $\teff$, \mh\space  and $\logg$ distributions are denoted on the plots. The error bars denote the mean statistical and total uncertainties for the stellar parameter measurements. A comparison with Field 0 stars (Fig.~4 of \citealt{kmoslt}) reveals that the mean values of the distributions are similar. The \teff\space distribution is slightly narrower (by 34\,K)  and cooler (by a mean \teff\space of 130\,K). The reason is probably that Field 0 contains a larger fraction of  fainter, slightly hotter stars. The metallicity distribution has a higher mean value of \mh=0.34\,dex than  Field 0 (0.26\,dex), and is slightly narrower ($\sigma_{\mh}$=0.39\,dex instead of 0.42\,dex in Field 0). The surface gravity has the largest uncertainties, and has also slightly lower values and a narrower distribution than Field 0. 

\subsection{Spatial   anisotropy for stars with \mh\textless0\,dex}
\label{sec:aniso}

We show the spatial distribution of the  stars in our data set  in Fig.~\ref{fig:metfield2} on the upper panel, the colours denote different \mh. Sub-solar metallicity stars with \mh\textless0.0\,dex are highlighted as square symbols. There are more sub-solar metallicity stars in the N and NW than in the S and SE. 

To confirm this finding and quantify the anisotropic distribution of sub-solar metallicity stars, we corrected  for the varying completeness (Table \ref{tab:sample}) of the data by applying a brightness cut to our sample. 
We only considered stars with an extinction corrected $K_{S,0}$\textless 10.25\,mag. Applying this cut allows us   to include the data from \cite{kmoslt}, which have a higher completeness than our data.  
We excluded rather young supergiant stars by considering only stars with $K_{S,0}$\textgreater 7.5\,mag. Our final   sample contains 729 stars, which are mostly red giant stars, and potentially asymptotic giant branch stars. The spatial distribution of these stars is shown in the middle panel of Fig.~\ref{fig:metfield2}. 
 
 To investigate spatial variations of the metallicity distribution, 
 we binned our  sample with a modified version of the Voronoi binning code of \cite{voronoi}. The original procedure performs spatial  binning of two-dimensional data such that each  bin is relatively round and has approximately the same signal-to-noise ratio, given a minimum S/N. 
 Our code distributes the stars such that we have approximately the same number of stars in a bin. We tried different realisations, with a different minimum number of stars (20, 25, and 30), and obtained consistent results. 
 
\begin{figure*}
    \includegraphics[width=0.95\textwidth]{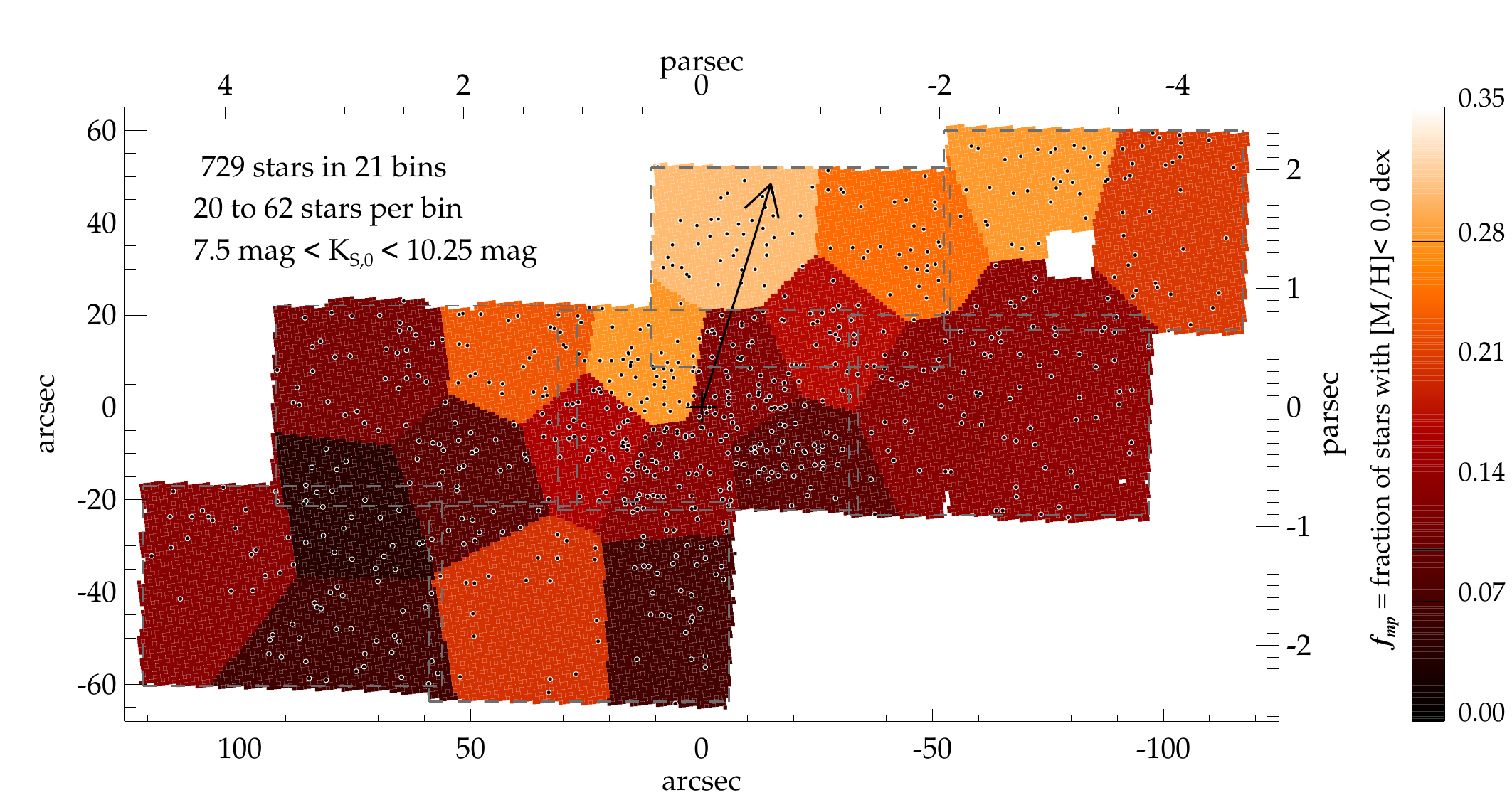}
 \caption{Binned map of the fraction of stars with \mh\textless0.0\,dex $f_\text{sub-s.m.}$ per bin, in Galactic coordinates offset from Sgr~A*. Bright orange colours denote a larger fraction of sub-solar metallicity stars. We also plot the stars in each bin as black points with white circles. The extent of the different fields is illustrated by grey dashed lines. The black arrow denotes the gradient of the fraction. The map was resampled such that one resolution element corresponds to four KMOS spatial pixels, i.e. 0.8\arcsec. The white field in the upper right is a hole caused by a technical problem during  observations. We find a  gradient of the fraction of sub-solar metallicity stars, with a larger fraction located in the Galactic North. }
 \label{fig:metfracvor}
\end{figure*}

 For each spatial bin, we calculated the fraction of stars with \mh\textless0.0\,dex, $f_\text{sub-s.m.}$. This fraction is more sensitive to the tail of sub-solar metallicity stars in the metallicity distribution than the mean or median metallicity.  We show a map of the sub-solar metallicity star fraction $f_\text{sub-s.m.}$ in Fig.~\ref{fig:metfracvor}. There is an increase of $f_\text{sub-s.m.}$ to the Galactic North and West, indicated by lighter colours. We also computed the gradient of the sub-solar metallicity star fraction,  indicated as black arrow in Fig.~\ref{fig:metfracvor}. In Galactic coordinates, the position angle of the metallicity-fraction gradient is at about 340\degr  (in equatorial coordinates 309\degr)   East of North, the slope is 2 per cent per 10 arcsec.  
 
 We tested the robustness of our metallicity-fraction gradient by applying several additional selection criteria to our sample of stars. In particular, we  excluded the stars for which we do not have coverage by the extinction map, located in the Galactic South at y\textless-50\arcsec\space in our maps. These stars were classified as member stars using  machine learning in Sec.~\ref{sec:selection}. Further, the extinction map may have systematic uncertainties in regions with high extinction, and underestimate the extinction in those regions. This can cause that foreground stars are considered as member stars. To detect such regions, we used $J$-band photometry, which is more affected by extinction than the $K$-band. We created a $J$-band number density map $n_J$ using the $J$-band catalog of stars by  \cite{exthawki}. Regions with low number  density $n_J$ of $J$-band sources indicate higher extinction. We excluded about 120 stars that are in regions where $n_J$ is less than the mode of the $n_J$ map. Both steps exclude stars in regions with rather high extinction. Our data set also contains stars with a rather low extinction. In our 2-layer extinction correction, we used a different extinction map for about 40 stars with observed 1.3\,mag\textless($H-K_S$)\textless1.7\,mag. We also tested excluding these stars from our sample.  All these cuts together reduce our sample from 729 to 562 stars, and the number of stars with \mh\textless0\,dex from 115 to 88.  Further, we tested a more stringent cut of $K_{S,0}$\textless 9.5\,mag instead of 10.25\,mag, which decreased the number of stars by 40\%. All these additional cuts and criteria do not  affect our main results. In all cases, the metallicity-fraction gradient points to  about 330\degr--350\degr\space East of North. The gradient of the median extinction however varies, depending on our sample of stars. This can be expected, as we excluded stars located in regions of high and low extinction. 
  
 Further, we  tested whether it is possible that the stars, randomly distributed over the observed field, produce the observed steepness of the metallicity-fraction gradient. Using our sample of 729 stars,   in 5000 runs we shuffled the values of \mh, calculated the $f_\text{sub-s.m.}$ in the same 21 bins as in Fig.~\ref{fig:metfracvor}, and measured the metallicity-fraction gradient. The median metallicity-fraction gradient of the 5000 runs is  0.52 percent per 10 arcsec.
Only  0.04 per cent, i.e. 2 out of 5000 runs, result in  a metallicity-fraction gradient of  $\ge$2 per cent per 10 arcsec, comparable to our data.  Thus, it is unlikely that our observation  is caused by statistical fluctuations. 
 
\begin{figure*}
  \includegraphics[width=0.95\textwidth]{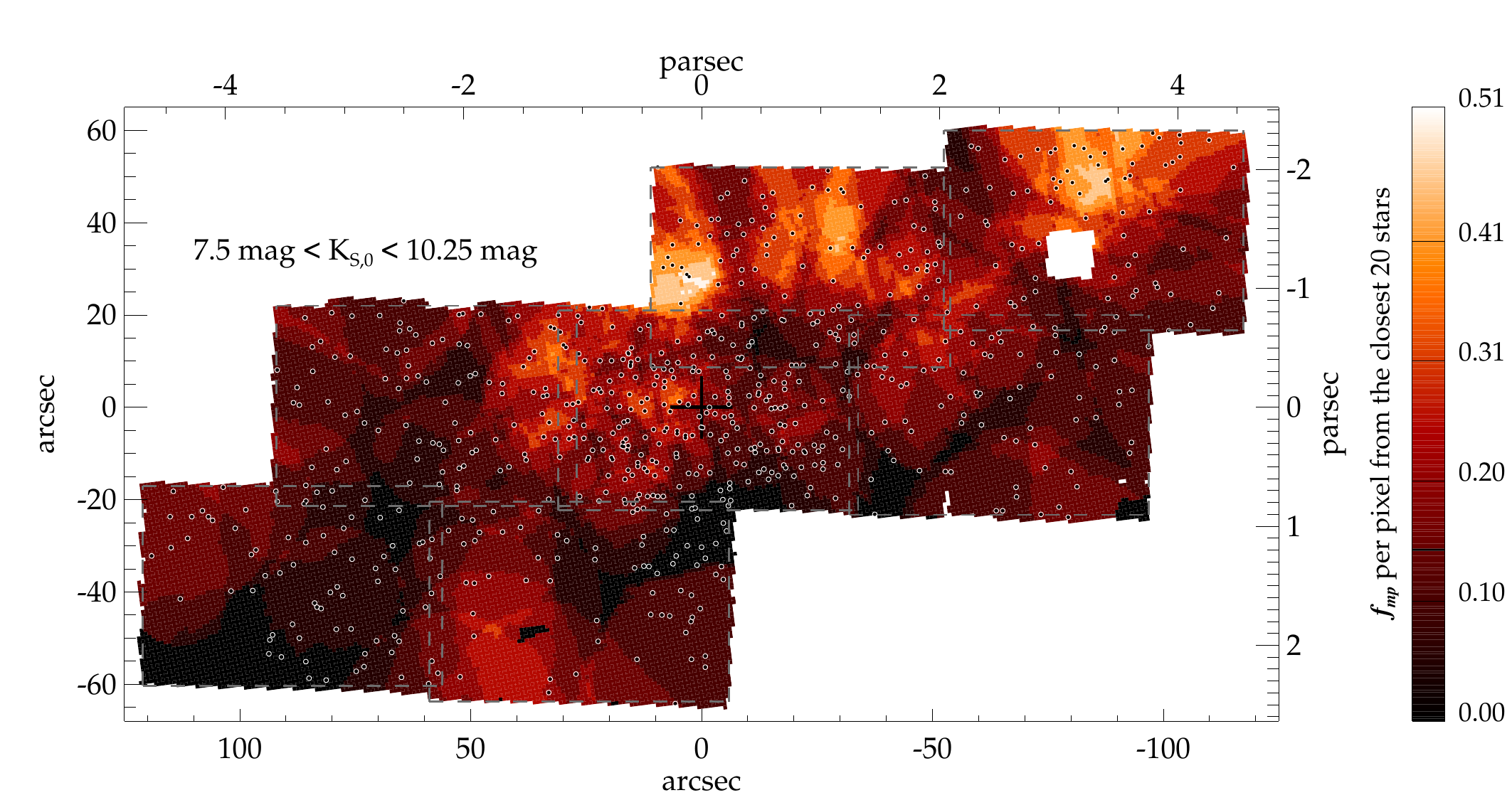}\caption{Map of the fraction of stars with \mh\textless 0.0\,dex, among the 20 closest stars in each pixel.   Coordinates, symbols and sampling as in Fig.~\ref{fig:metfracvor}, however, the colour bar has a different range.  The fraction of sub-solar metallicity stars is increasing towards  the North of our field. }
  \label{fig:metfrac20}
\end{figure*}

 For a finer spatial resolution, we searched for the 20 closest stars of each 0.2\arcsec$\times$0.2\arcsec\space pixel in our field.  From these 20 stars we computed the fraction of stars with \mh\textless0.0\,dex, $f_\text{sub-s.m.}$. The result is shown in Fig.~\ref{fig:metfrac20}, adjacent pixels are  correlated, and the spatial resolution depends on the stellar number density in a given region. Nevertheless, the resolution is finer than in Fig.~\ref{fig:metfracvor}.   The general appearance of the maps is similar, with higher fractions of sub-solar metallicity stars in the North. This confirms that the metallicity distribution variation is not caused by spatial binning.  

\subsection{Metallicity and radial velocity distributions in different regions of the nuclear star cluster}
\label{sec:metdistvgl}

We found an asymmetry in the distribution of sub-solar metallicity stars, with a larger fraction of \mh\textless0\,dex in the Galactic North and West of the nuclear star cluster compared to the South and East. 
In this section we investigate whether the change of the sub-solar metallicity star fraction is due to a global shift of the metallicity distribution, or caused by a low-metallicity tail in the metallicity distribution. 

\begin{figure}
    \centering  
  \includegraphics[width=\columnwidth]{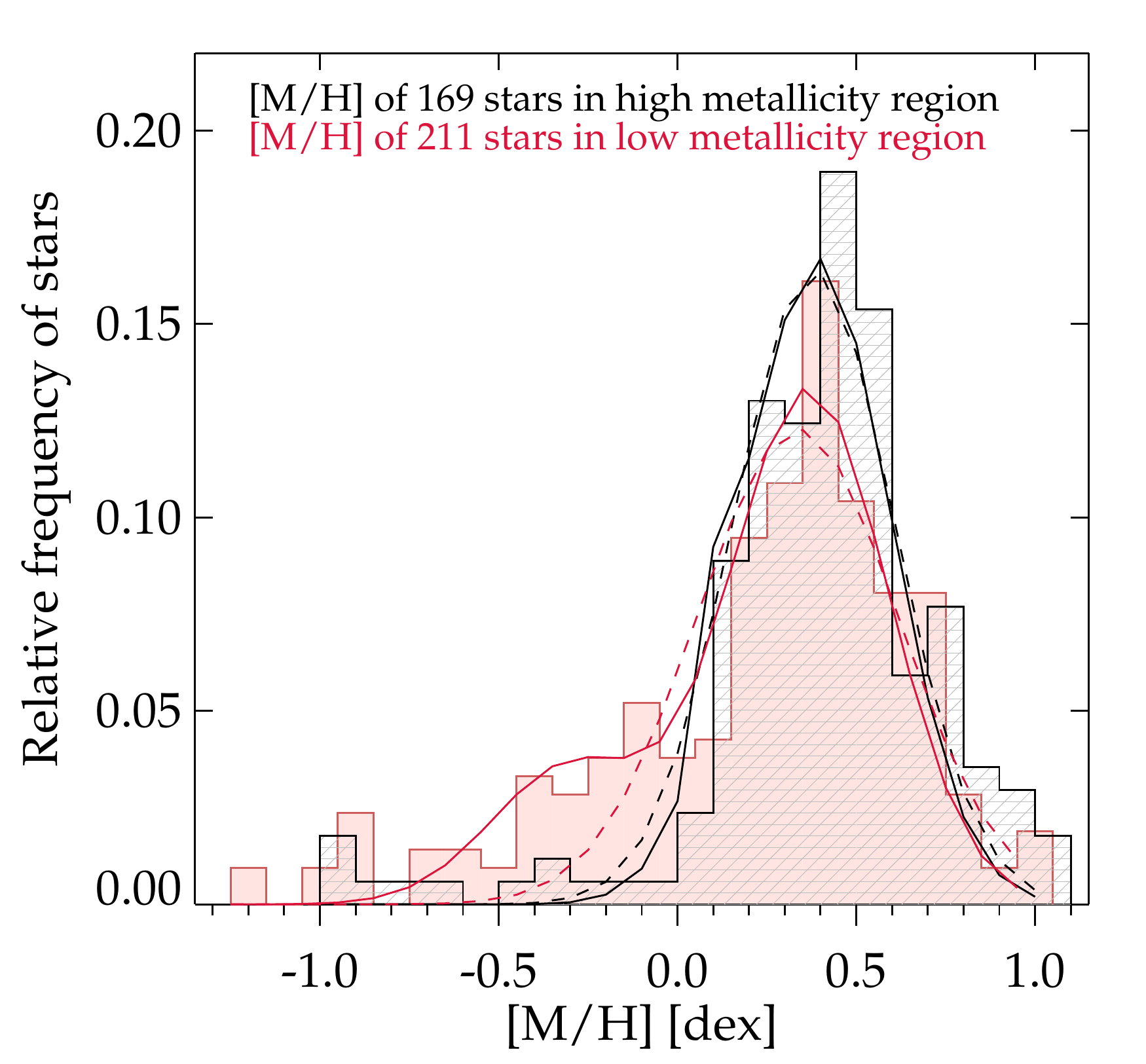}
\caption{Normalised metallicity distribution of stars in the  high-metallicity region (black colour, fraction of sub-solar metallicity stars $f_\text{sub-s.m.}$\textless 10 per cent) and low-metallicity region (red, fraction of sub-solar metallicity stars $f_\text{sub-s.m.}$\textgreater 20 per cent). The black dashed line denotes a Gaussian fit to the metallicity distribution of the high-metallicity region; the red dashed line  of the low-metallicity region; the red solid line a double-Gaussian fit of the low-metallicity region. The double-Gaussian fit is required to fit the tail of sub-solar metallicity stars.}
 \label{fig:histfrac}
\end{figure}

We selected stars in the Voronoi bins (Fig.~\ref{fig:metfracvor}) with  $f_\text{sub-s.m.}$ \textgreater 20\% as low-metallicity group, and stars in the bins with $f_\text{sub-s.m.}$ \textless 10\% as high-metallicity group. The two groups contain 211 and 169 stars. 
We show the two different normalised metallicity distributions  in Fig.~\ref{fig:histfrac}. The metallicity distribution in the region with a higher fraction of sub-solar metallicity stars has a low-metallicity tail at \mh\textless 0.0\,dex. We fitted a Gaussian function to the metallicity distributions.  In the high-metallicity  region, the  Gaussian is located at \mh=0.39\,dex with $\sigma$=0.3\,dex. This distribution is reasonably  well represented by a Gaussian function. For the metallicity distribution in the low-metallicity region, we obtained a Gaussian located at \mh=0.34\,dex with $\sigma$=0.4\,dex. However, the tail of sub-solar metallicity stars with \mh\textless0.0\,dex led us to perform a double Gaussian fit to the metallicity distribution in the low-metallicity region. The higher  peak is located at 0.37\,dex with $\sigma$=0.3\,dex; the second, low-metallicity peak  at \mh=--0.29\,dex, and $\sigma$=0.3\,dex. The exact results of the Gaussian fits depend on the binning of the histograms. But irrespective of the binning, the metallicity distribution in the low-metallicity region is better described by a double-Gaussian distribution than a single Gaussian.

The histograms indicate that there are metal-rich populations of stars in both regions, with similar Gaussian distributions, located at \mh$\approx$0.38\,dex with $\sigma\approx$0.3\,dex. In both regions, there are also sub-solar metallicity stars. 
However,  the  Galactic North-Western region of the nuclear star cluster contains   a larger relative frequency sub-solar metallicity stars. 
The nuclear star cluster's stellar populations are not homogeneous and not isotropic around Sgr A*. 

\begin{figure}
    \centering  
  \includegraphics[width=\columnwidth]{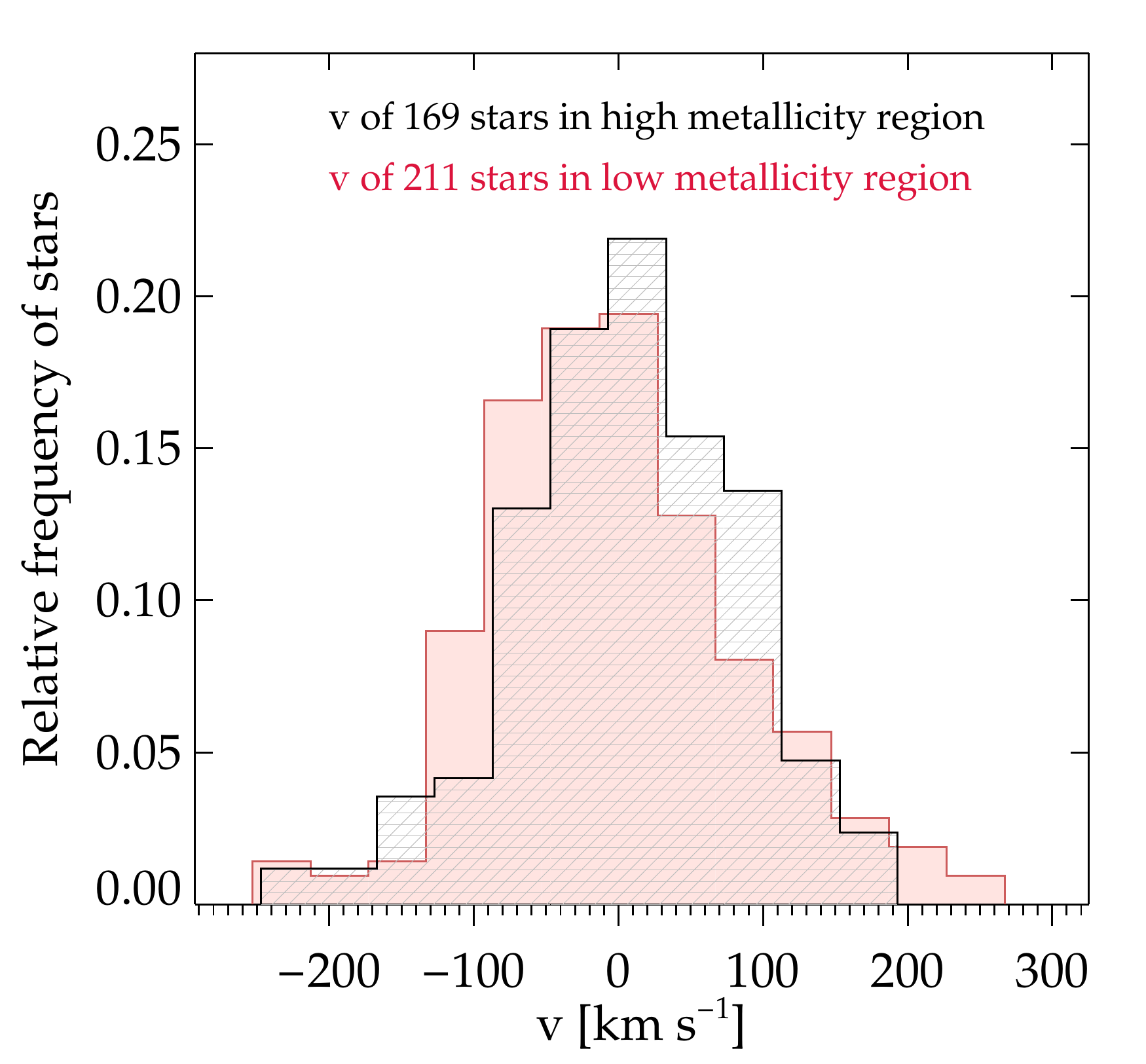} 
  \caption{Normalised radial velocity distribution of stars in the  high-metallicity region (black colour, fraction of sub-solar metallicity stars $f_\text{sub-s.m.}$\textless 10 per cent) and low-metallicity region (red, fraction of sub-solar metallicity stars $f_\text{sub-s.m.}$\textgreater 20 per cent).}
 \label{fig:histfracrv}
\end{figure}

We also plot histograms of the radial velocity of stars in the high-metallicity and low-metallicity regions in Fig.~\ref{fig:histfracrv}. The mean velocities differ by  10\,\kms, which may be due to the different locations of the regions and the rotation of the nuclear star cluster. The velocity dispersion in the low-metallicity region is higher by 9\,\kms. Although the mean projected distances of the stars in the two regions are similar ($\sim$55 arcsec), we cannot exclude that the  velocity dispersion difference is caused by the spatial distribution of the stars. A more detailed kinematic analysis is required, but this is  beyond the scope of this paper. We publish the radial velocity measurements with our stellar parameters online.

\subsection{A low fraction of  metal-poor stars   with  \mh\textless--0.5\,dex }

So far we consider sub-solar metallicity stars with \mh\textless0\,dex. We chose this cut  because we found that the  metallicity distribution in different spatial regions varies at  \mh\textless0.0\,dex (see Sec.~\ref{sec:metdistvgl}). 
Other publications considered metal-poor stars in the Galactic center as stars with \mh\textless--0.5\,dex, and computed the metal-poor star fraction. In order to enable a comparison with the literature, we use the criterion  \mh\textless--0.5\,dex for metal-poor stars in this section. We note that this definition deviates from the classification suggested by \cite{2005ARA&A..43..531B}, where metal-poor stars have --2.0\,dex\textless[Fe/H]$\leq$--1.0\,dex.

In the entire area covered  by Fields 1-6, we obtain $f_{mp}$=3.5 per cent.  \cite{dolowfe} found 6 per cent of their stars located at projected radii r\textless 0.5\,pc  have \mh\textless--0.5\,dex;  \cite{kmoslt} obtained a similar value in the central Field 0 with r\textless 1.4\,pc, 5.2 per cent. 
We tested if the  lower fraction in our data compared to Field 0 in \cite{kmoslt} is  due to our lower completeness rather than the different spatial coverage. 
We made magnitude cuts and considered only stars with 7.5\,mag\textless $K_{S,0}$\textless 10.25\,mag. After this cut, the fractions of stars with \mh\textless--0.5\,dex change only little, and the discrepancy between Field 0 and the combined Fields 1-6 remains. 
We also tested if there are spatial variations of $f_{mp}$ with \mh\textless--0.5\,dex in Fields 1-6, but since the total number of stars with \mh\textless--0.5\,dex is only 20, we are more sensitive to binning and data selection effects, therefore the following results need to be considered with care. 
We found that  $f_{mp}$  has a much shallower gradient than for \mh\textless0\,dex, changing only by about 0.2 per cent per 10 arcsec instead of 2 per cent per 10 arcsec. The direction  of the gradient is in agreement  with the gradient for \mh\textless 0.0\,dex, with a larger $f_{mp}$ towards the North, at about 340\degr\space East of North. 

\section{Discussion}
\label{sec:sec5}

\subsection{Metallicity distributions in the literature}
In agreement with our previous work \citep{kmoslt}, where we studied stars in the nuclear star cluster out to 1.4\,pc, we found that the majority of stars are metal-rich. Also \cite{dolowfe} found a large fraction of metal-rich stars in the central 1\,pc of the Galactic centre, and in addition stars with \mh$\lesssim$0\,dex. 

\cite{rydechem14} measured [Fe/H] for 9 M giants in the Galactic centre, but at larger projected distances from  Sgr A* than our data. Their data set was metal-rich, with a  mean [Fe/H]=0.11$\pm$0.15\,dex. The data were reanalysed by \cite{2018MNRAS.478.4374N}, who found an even higher mean [Fe/H] of 0.3\,dex.  
\cite{2017AJ....154..239R}  obtained the so far largest sample at  high spectral resolution ($R$ $\sim$24,000), with 17 M giants in the nuclear star cluster and nuclear disk. 
They  obtained a mean iron-based metallicity of [Fe/H]=--0.11\,dex (median [Fe/H]=--0.16\,dex) for their 17 stars, ranging from $-1.15$ to +0.64\,dex.  This is a lower mean value than we obtain, however, there are several differences in our sample and analysis.   \cite{2017AJ....154..239R} used the iron-based metallicity [Fe/H], while we used the overall metallicty \mh, meaning that all elements are considered, not only Fe.  Further, we have different assumptions on [$\alpha$/Fe]. \cite{2017AJ....154..239R} assumed  [$\alpha$/Fe]=0.3\,dex for stars with [Fe/H]=--0.5\,dex, with  [$\alpha$/Fe] decreasing linearly with increasing [Fe/H] up to [Fe/H]=0\,dex, and [$\alpha$/Fe]=0\,dex at [Fe/H]\textgreater 0\,dex, whereas we assumed that [$\alpha$/Fe]=0\,dex at all values of [M/H]. This  causes differences for stars with subsolar metallicity.  Concerning the sample of stars, the median $K_{S}$ magnitude of \cite{2017AJ....154..239R} is 1 mag brighter than our median $K_{S}$. 

For  few giant stars in the nuclear star cluster a detailed abundance analysis was performed so far.   \cite{2016ApJ...831...40R} studied a high-resolution spectrum ($R$ $\sim$24,000) of a metal-poor giant star in the Galactic centre, and found [Fe/H]$\approx$--1.0\,dex and [$\alpha$/Fe]$\approx$0.4\,dex, which confirms  the presence of metal-poor giant stars in the Galactic centre region with high spectral resolution data.  \cite{2018ApJ...855L...5D} investigated the other extreme of the metallicity distribution, and observed  metal-rich stars of the nuclear star cluster with high spectral resolution ($R$ $\sim$25,000).  They confirmed the high value of the overall metallicity \mh\space found with medium-resolution data \citep{dolowfe} for one of the stars. But they also found that model spectra  cannot reproduce all features of metal-rich stars. This may affect also  the accuracy of our  results at super-solar metallicities. As noted in Sect. \ref{sec:unc}, the systematic uncertainties for metal-rich stars \mh\textgreater0.3\,dex may be underestimated. 

Our fraction of metal-rich stars with \mh$\geq$0.3\,dex  $f_{mr}$ is 0.6.  \cite{2018MNRAS.478.4374N} obtained $f_{mr}$=0.4, and \cite{2017AJ....154..239R} $f_{mr}$=0.2. \cite{dolowfe}, who have a more concentrated sample located within 1 pc of the nuclear star cluster, found  a higher fraction of $f_{mr}$=0.7.

In summary, our larger data set  confirms what has been found in  previous studies with smaller samples: The nuclear star cluster  has a high fraction of metal-rich stars \mh\textgreater 0\,dex, but also a non-negligible number of sub-solar metallicity stars. Differences to other studies are caused by different assumptions, methods, and samples.

\subsection{Extinction  and completeness effects}
\begin{figure*}
    \includegraphics[width=0.95\textwidth]{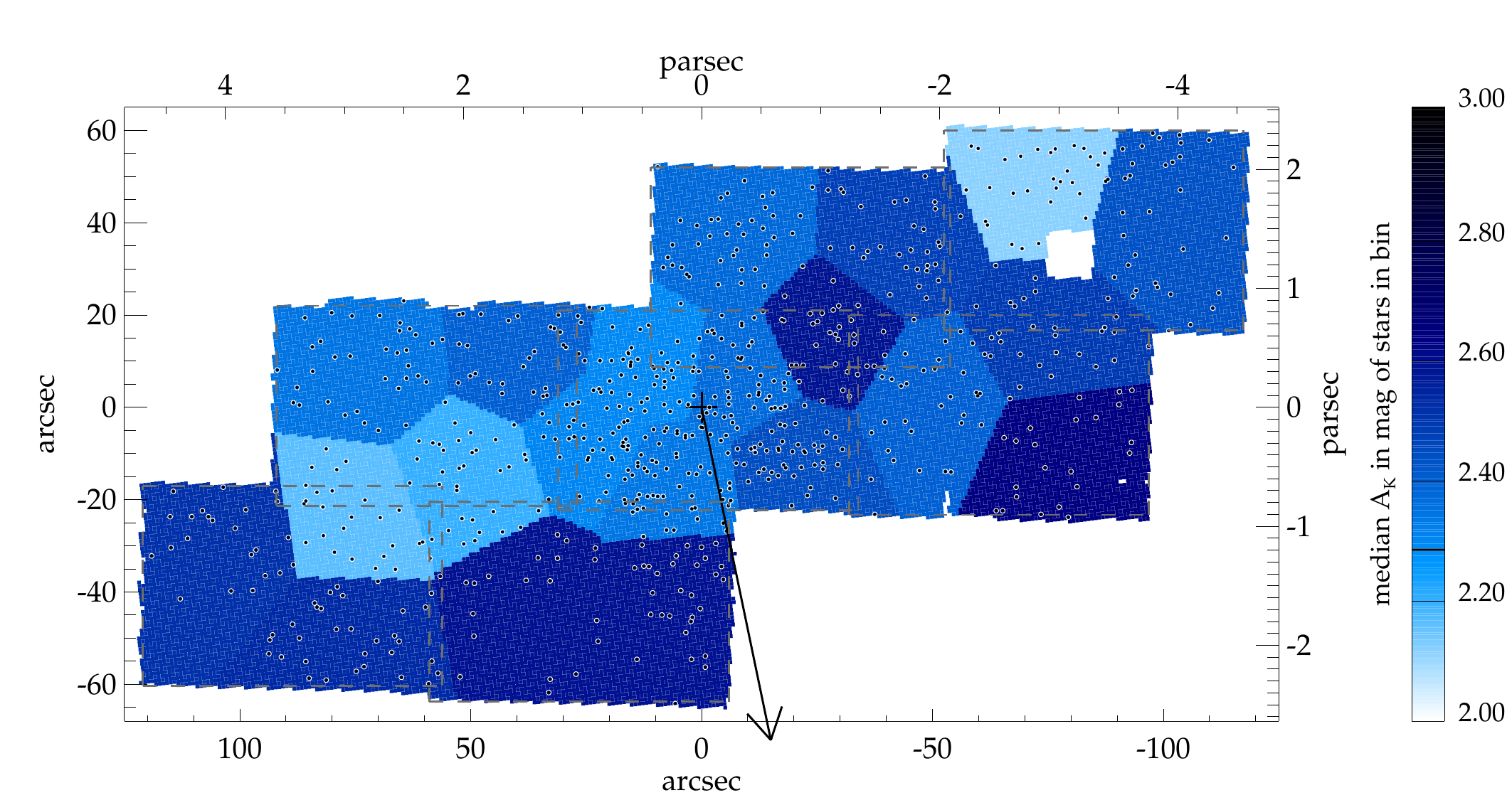}
 \caption{Binned map of the median extinction $A_K$ of the stars in each bin. Darker  colours denote a higher median extinction.  The black arrow denotes the gradient of the median extinction.   Coordinates, symbols and sampling as in Fig.~\ref{fig:metfracvor}. The median extinction per bin is not correlated with the fraction of sub-solar metallicity stars.}
 \label{fig:metakvor}
\end{figure*}
\begin{figure}
    \centering  
  \includegraphics[width=\columnwidth]{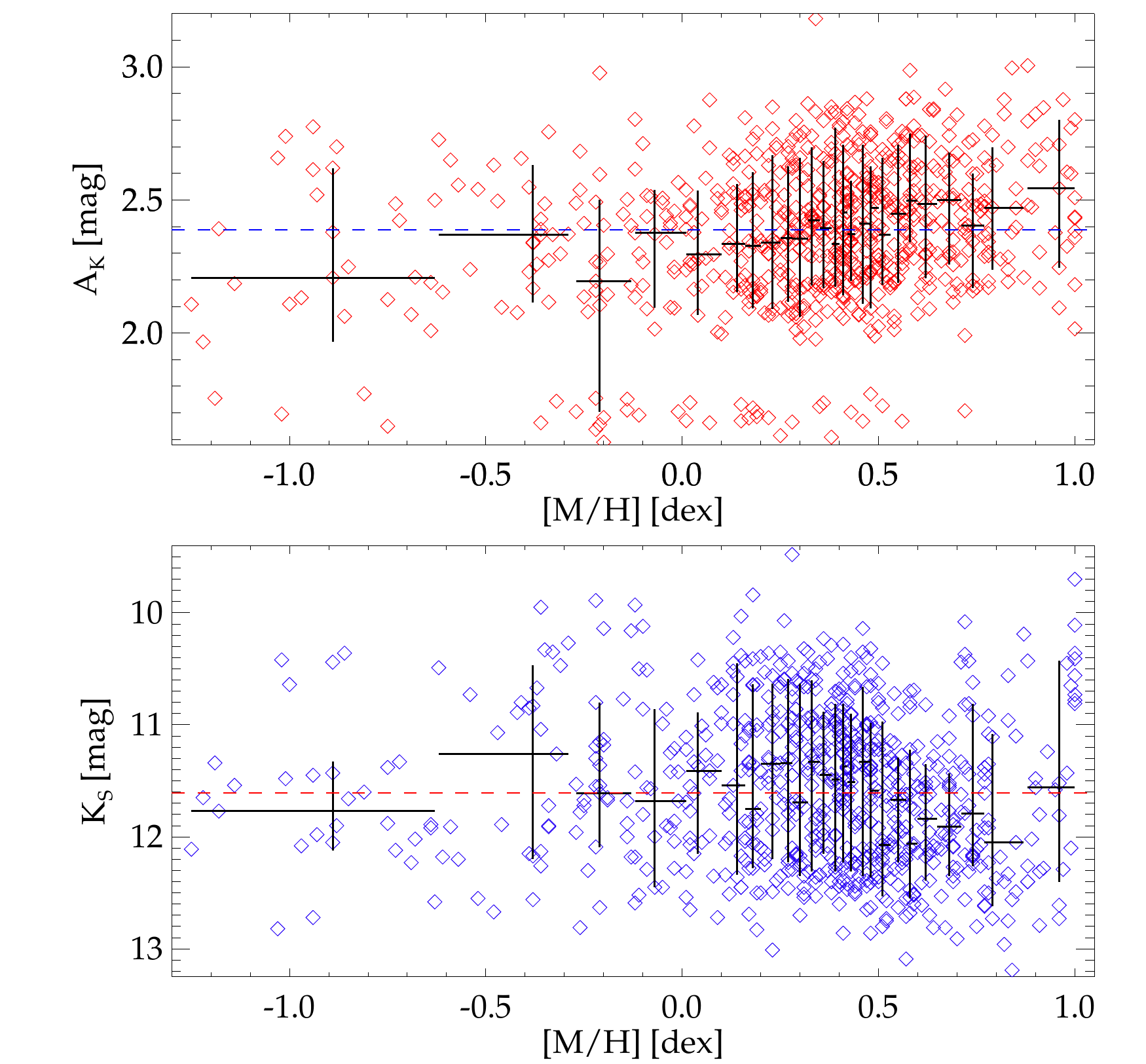}
 \caption{Stellar metallicity of 729 stars plotted against their local extinction $A_K$ (upper panel) and observed $K_S$ magnitude (lower panel). The horizontal dashed lines denote the overall median values of $A_K$  and $K_S$, while the black data points denote the median values of  $A_K$  and $K_S$ for 30 stars,  sorted by \mh. The horizontal black lines show which range of \mh\space a given bin spans, the vertical black lines denote the 33. and 67. percentiles. 
}
 \label{fig:binnedmh}
\end{figure}

Extinction in the Galactic centre is high and variable, with  $A_K$ ranging from $\lesssim$1.6 to $\gtrsim$ 3.2\,mag. Local changes in the extinction can mean that our stellar spectra lie deeper within the nuclear star cluster, or mostly in the outer regions. We cannot say for sure where a given star is, as we do not know individual distances. However, by looking at the median extinction in our Voronoi bins, we can at least test if a region has higher extinction, and  a dark cloud along the line-of-sight may prevent a deeper look into the nuclear star cluster. We calculated the median extinction of the stars in a Voronoi bin with the extinction maps of \cite{exthawki}. The results are shown in Fig.~\ref{fig:metakvor}, with the same binning as in Fig.~\ref{fig:metfracvor}.  The median extinction values range from 2.1 to 2.6 mag, but are not correlated with the fraction of sub-solar metallicity stars shown in Fig.~\ref{fig:metfracvor}. We also computed a gradient, shown as black arrow in  Fig.~\ref{fig:metakvor}, and found that it  is approximately 150\degr\space offset to the metallicity-fraction gradient. The gradient changes with binning,  within a range of 45\degr. The reason is that extinction varies on smaller scales than the size of our Voronoi bins. 

We also analysed the stellar metallicity \mh\space as a function of the extinction $A_K$  for each star individually (Fig.~\ref{fig:binnedmh}, upper panel). 
The values for $A_K$ have a gap, with few stars having $A_K$ between 1.8 and 2.0\,mag. This is due to our 2-layer extinction correction (Sec. \ref{sec:selection}), where we applied different extinction maps for stars with different observed colour $H-K_S$. We also plot the median extinction for 30 stars, sorted by their value of \mh, as black crosses. The vertical error bars denotes the 33. and 67. percentile of the $A_K$ distributions. All bins lie well within the median extinction of all stars (horizontal dashed line). There may be a mild correlation between  \mh\space and $A_K$, such that stars with lower \mh\space are in regions with lower $A_K$, but the variation of the median $A_K$ is only 0.35\,mag. A linear fit to the median $A_K$ as a function of \mh\space has a slope of 0.16$\pm$0.13\,mag$\cdot$dex$^{-1}$. When we apply  additional cuts to remove stars in regions with high and low extinction regions, as in Sec.~\ref{sec:aniso}, the relation disappears and the slope is 0.07$\pm$0.1\,mag$\cdot$dex$^{-1}$. This shows that there is no significant correlation of \mh\space and $A_K$, 
and  our results of the spatial anisotropy of sub-solar metallicity stars are not caused by  variable extinction. 

Likewise, the varying completeness over the field-of-view of our data is unlikely to cause the change in the fraction of sub-solar metallicity stars $f_\text{sub-s.m.}$. We made magnitude cuts to ensure that the faint population of stars, which is distributed unevenly  among fields due to varying completeness, do not bias our results. All fields contain stars in the same magnitude range.  Also, the Fields 1 and 2 in the North and North-West, which have a high  fraction of sub-solar metallicity stars, have very different completeness (see Table \ref{tab:sample}), they are the fields with the  lowest and highest   completeness (with exception of the central field). Yet,  both fields contain a higher fraction of sub-solar metallicity stars than other fields, with completeness values in between. We conclude that varying completeness does not cause the variation of the sub-solar metallicity star fraction.

\subsection{Data selection effects}

We tested whether we introduce any bias in the distribution of sub-solar metallicity stars when we deselect stars with low S/N and bad fit quality.
First, we tested if the sub-solar metallicity stars are significantly fainter than metal-rich stars, which would suggest that they are more likely to be deselected. We  show the stellar metallicity \mh\space as a function of the observed $K$-band magnitude $K_S$  (Fig.~\ref{fig:binnedmh}, lower panel). We found no  correlation of \mh\space with $K_S$, a linear fit to the median $K_S$ as a function of \mh\space gives a slope of 0.06$\pm$0.29\,mag$\cdot$dex$^{-1}$.  This means that the sub-solar metallicity stars are not significantly brighter or fainter than metal-rich stars in our sample, an effect that might be caused by a biased sample selection.

Further, we investigated the spectra that were deselected for our analysis. For each field, we stacked spectra of stars with 5\textless S/N \textless20, and applied full spectral fitting to the six stacked spectra. If the stacked spectra in the  fields 4-6 had a lower metallicity than the spectra in the fields 1-2, this would suggest that we deselected sub-solar metallicity stars in the South or metal-rich stars in the North, and introduced a bias. Before stacking the spectra, we used their radial velocities measured in Sec.~\ref{sec:sec3} with \textsc{pPXF} to shift them to rest wavelength.  We summarise our results in Table \ref{tab:stack}. The number of stars per stack varies from 27 to 96, but most fields have 47 to 68 stars that we used for stacking. The median of the extinction-corrected $K_{S,0}$ magnitude of the stacked stars  is fainter than our sample of stars, which we constrained to $K_{S,0}$\textless 10.25\,mag. The stacked spectra in all six fields have supersolar metallicity, which confirms that the nuclear star cluster is metal-rich, also for fainter stars. While there is some variation of the resulting metallicity for the six fields, we find no trend to lower metallicities in Fields 4-6 compared to Fields 1-2, which suggests that we do not introduce a bias to the fraction of sub-solar metallicity stars when we perform out data selection. 
\begin{table}
 \centering
 \caption{Summary of stacked spectra}
 \label{tab:stack}
\begin{tabular}{@{}cccccc@{}}
\noalign{\smallskip}
\hline
\noalign{\smallskip}
Field& Number of & median $K_{S,0}$ &\mh& $\sigma_\text{fit}$&$\sigma_\text{fit}$\\
&	stacked spectra &[mag]&[dex]&[dex]&[dex]\\ \noalign{\smallskip}
\hline
\noalign{\smallskip}
1	&62&11.9 &0.07&+0.03 &-0.02 \\
2	&96&12.5 &0.13&+0.03 &-0.02\\
3	&47&11.7 &0.18&+0.02 &-0.02 \\
4	&68&11.3 &0.24&+0.02 &-0.02\\
5	&60&12.4 &0.13&+0.02 &-0.02\\
6	&27&10.6 &0.31 &+0.01 &-0.02\\
 \hline  
 \end{tabular}
\end{table}


\subsection{Possible origin of the metallicity asymmetry}
The formation of nuclear star clusters is still under debate. It has been proposed that (a) the stars formed  `in situ', i.e. in the Galactic centre \citep[e.g.][]{milos04,pflamm09}; or (b) a `wet merger' scenario, where massive star clusters formed in the Galactic disk, and migrated to the centre while continuing to form stars from their gas reservoir \citep{2016MNRAS.461.3620G};  or  (c)  a `dry merger' scenario, where star clusters formed  `ex situ',   migrated to the Galactic centre through dynamical friction, and merged to the nuclear star cluster \citep[e.g.][]{tremaine75,antonini12,2015ApJ...806..220A,2018MNRAS.477.4423A}. These scenarios are able to produce the observed  mixed stellar populations \citep{perets14,2015ApJ...799..185A,2016MNRAS.461.3620G}, and thus also a  broad metallicity distribution. 

The spatial anisotropy of sub-solar metallicity stars may indicate that some of them were brought to the Galactic centre from  star cluster infall events. \cite{perets14} studied the distribution of stellar populations  in $N$-body simulations of repeated star cluster  infall events. The infalling star clusters resemble massive globular clusters in their density distribution, and started at an orbital radius  of 20\,pc \citep{antonini12}.  The simulation shows   that the different  stellar populations originating from  the star clusters have distinct three-dimensional structures, and some structures are highly anisotropic even Gyr after their infall. Similar simulations were performed in \cite{2018MNRAS.479..900A}, and they found that the initial spatial distribution is  determined by the orbit of the infalling star cluster.

The infall time of the star clusters depends on the mass and distance to the nuclear star cluster. For example, a cluster with  mass 10$^5$--10$^7$\,M$_{\sun}$ and starting at  $r\sim$2--5\,kpc  could have reached the Galactic centre \textless3\,Gyr ago, just as an infalling dwarf galaxy with initially a few 10$^8$ to 10$^{10}$\,M$_{\sun}$, and starting at $r$ between a  few ten to a few hundred kpc \citep{mas}.  
  The Milky Way has several globular star clusters located in the Galactic  bulge.   Within a galactic-centric radius of 2\,kpc, almost 50 per cent of the clusters have [Fe/H]\textgreater$-1.0$\,dex, and about 20 per cent even  [Fe/H]\textgreater$-0.5$\,dex \citep{harris,2010arXiv1012.3224H}.  However, the census of Galactic globular  clusters is not yet  complete, and new clusters were discovered recently \citep{2018ApJ...860L..27C,2019MNRAS.484L..90C}. 

More information is required to determine the origin of  sub-solar metallicity  stars in the nuclear star cluster. If some of them originate from a star cluster infall, they should have the same metallicity and element abundances.  Our metallicity measurements have  large uncertainties of $\sigma_{\mh}\approx$\,0.26\,dex, which is larger than  the internal metallicity dispersion of Galactic globular clusters. We require high-resolution spectroscopy, and precise 
element abundance measurements to confirm the hypothesis that  a star cluster infall caused the spatial anisotropy of sub-solar metallicity stars.  If several infall events happened, element abundance measurements  may be able to separate the different stellar populations further, and their common chemistry will show which stars likely formed together.

Another way to investigate the origin of the metallicity asymmetry is to combine the metallicities with kinematic measurements, i.e. radial velocities and proper motions. If indeed a star cluster or dwarf galaxy infall to the nuclear star cluster happened not longer ago than the relaxation time,  the population  can be distinguished from its kinematic properties, as shown in \cite{mas} using $N$-body simulations. Distinct kinematics were indeed found by \cite{tuanprep} for  sub-solar metallicity stars located in  the central Field 0. 
Future analysis can reveal if also the sub-solar metallicity stars found in this study show  distinct kinematics from the super-solar metallicity stars. To enable such an analysis, we publish the radial velocity measurements with our stellar parameters online.

\section{Conclusions}
We observed almost half of the area of the Milky Way's nuclear star cluster with the integral-field spectrograph KMOS. We extracted $K$-band spectra of more than 600 late-type stars, and derived stellar parameters using full-spectral fitting. Most stars are red giant stars, with metallicities ranging from \mh=--1.25\,dex to \textgreater +0.3\,dex. We investigated the spatial distribution of sub-solar metallicity stars with \mh\textless0.0\,dex. The Galactic North and North-West region of our observed field has a more than two times larger fraction of sub-solar metallicity stars than the region in the  Galactic South-East. A comparison of the metallicity histograms in the two regions revealed a tail of stars with \mh\textless0.0\,dex  in the  low-metallicity region. One possible explanation for such an  anisotropic metallicity distribution is  a recent merger event of a sub-solar metallicity stellar population, which has not yet mixed completely with the more metal-rich stars of  the nuclear star cluster. 
 \label{sec:sec6}
 \section*{Acknowledgments}

We would like to thank the ESO staff who helped us to prepare our observations and obtain the data.   We are grateful to   Lodovico Coccato  and Yves Jung for advice and assistance in the data reduction process. We thank Barbara Lanzoni for sharing her SINFONI data of NGC~6388. We also thank the referee for useful comments and suggestions.

N. N. and F. N.-L. gratefully acknowledge funding by the Deutsche Forschungsgemeinschaft (DFG, German Research Foundation) -- Project-ID 138713538 -- SFB 881 (``The Milky Way System'', subproject B8).
The research leading to these results has received funding from the European Research Council under the European Union's Seventh Framework Programme (FP7/2007-2013) / ERC grant agreement n. [614922] (RS and FNL). RS and FNL acknowledge
  financial support from the State Agency for Research of the Spanish
  MCIU through the "Center of Excellence Severo Ochoa" award for the
  Instituto de Astrof\'isica de Andaluc\'ia (SEV-2017-0709). RS acknowledges financial support from national project
  PGC2018-095049-B-C21 (MCIU/AEI/FEDER, UE).  
  ACS acknowledges financial support from NSF grant AST-1350389
This research made use of the SIMBAD database (operated at CDS, Strasbourg, France).  This research made use of Montage. It is funded by the National Science Foundation under Grant Number ACI-1440620, and was previously funded by the National Aeronautics and Space Administration's Earth Science Technology Office, Computation Technologies Project, under Cooperative Agreement Number NCC5-626 between NASA and the California Institute of Technology.

\footnotesize{
\bibliography{bibs_lt}
}
\appendix

\section{Table of stellar parameters}
\label{sec:tabsec}
\onecolumn
\begin{table}
\caption{Stellar parameters: Stellar identification number Id, the coordinates in R.A. and Dec, the stellar parameters $\teff$, $\mh$, $\logg$, $v_z$, and extinction corrected $K_S$-band magnitude $K_{S,0}$.  The full table is available online.}
\label{tab:parameter}
\begin{tabular}{ccccccccccccc}
Id & R.A.& Dec & $\teff$ & $\sigma_{\teff}$ & $\mh$&$\sigma_{\mh}$ & $\logg$&$\sigma_{\logg}$ & $v_z$&$\sigma_{v_z}$& $K_{S,0}$\\
&$ \left( ^{\circ} \right)$&$ \left( ^{\circ} \right)$&$ \left( \mathrm{K} \right)$&$ \left( \mathrm{K} \right)$&$ \left( \mathrm{dex} \right)$&$ \left( \mathrm{dex} \right)$&$ \left( \mathrm{dex} \right)$&$ \left( \mathrm{dex} \right)$&$\left( \mathrm{km}\,\mathrm{s}^{-1} \right)$&$\left( \mathrm{km}\,\mathrm{s}^{-1} \right)$&$\left( \mathrm{mag} \right)$&\\
\hline
$1000002$&$  266.41144 $&$   -29.003521 $&$  3458 $&$ ^{+305}_{-305} $&$ -0.36 $&$ ^{+0.25}_{-0.25} $&$  0.5 $&$ ^{+1.2}_{-1.2} $&$     43.1 $&$ ^{+ 7.8}_{- 7.8} $&$   8.29 $ \\
\\
$1000054$&$  266.41284 $&$   -29.005566 $&$  3507 $&$ ^{+209}_{-209} $&$  0.10 $&$ ^{+0.24}_{-0.24} $&$  0.0 $&$ ^{+1.0}_{-1.0} $&$    -29.0 $&$ ^{+ 3.4}_{- 3.4} $&$   8.53 $ \\
\\
$1000112$&$  266.40930 $&$   -29.002455 $&$  3576 $&$ ^{+207}_{-207} $&$ -0.36 $&$ ^{+0.25}_{-0.25} $&$  0.1 $&$ ^{+1.0}_{-1.0} $&$     41.1 $&$ ^{+ 0.9}_{- 0.9} $&$   8.67 $ \\
\\
$1000123$&$  266.40680 $&$   -29.002874 $&$  3744 $&$ ^{+255}_{-255} $&$  0.22 $&$ ^{+0.30}_{-0.30} $&$  0.2 $&$ ^{+1.0}_{-1.0} $&$     90.0 $&$ ^{+ 6.3}_{- 6.3} $&$  11.29 $ \\
\\
$1000149$&$  266.40729 $&$   -29.002579 $&$  3259 $&$ ^{+252}_{-252} $&$  0.29 $&$ ^{+0.41}_{-0.41} $&$  1.2 $&$ ^{+1.0}_{-1.0} $&$     53.7 $&$ ^{+ 4.6}_{- 4.6} $&$   8.51 $ \\
\\
$1000183$&$  266.41409 $&$   -29.005821 $&$  3373 $&$ ^{+206}_{-206} $&$  0.33 $&$ ^{+0.28}_{-0.28} $&$  0.1 $&$ ^{+1.0}_{-1.0} $&$     15.8 $&$ ^{+ 2.3}_{- 2.3} $&$   8.56 $ \\
\\
$1000196$&$  266.41214 $&$   -29.005774 $&$  3123 $&$ ^{+244}_{-244} $&$ -0.39 $&$ ^{+0.25}_{-0.25} $&$  0.6 $&$ ^{+1.1}_{-1.1} $&$    121.0 $&$ ^{+ 1.4}_{- 1.4} $&$   8.30 $ \\
\\
$1000216$&$  266.40958 $&$   -29.003115 $&$  3036 $&$ ^{+211}_{-211} $&$  0.18 $&$ ^{+0.27}_{-0.27} $&$  0.8 $&$ ^{+1.0}_{-1.0} $&$     17.4 $&$ ^{+ 2.4}_{- 2.4} $&$   8.13 $ \\
\\
\hline
\end{tabular}
\end{table}
\bsp

\label{lastpage}

\end{document}